# Happy New Year *Homo erectus*? More evidence for interbreeding with archaics predating the modern human/Neanderthal split.

**Peter J. Waddell**[1]

**pwaddell.new@gmail.com**

[1]Ronin Institute, 19 Flowermound Dr, West Lafayette, IN 47906, U.S.A.

A range of a priori hypotheses about the evolution of modern and archaic genomes are further evaluated and tested. In addition to the well-known splits/introgressions involving Neanderthal genes into out-of-Africa people, or Denisovan genes into Oceanians, a further series of archaic splits and hypotheses proposed in Waddell et al. (2011) are considered in detail. These include signals of Denisovans with something markedly more archaic and possibly something more archaic into Papuans as well. These are compared and contrasted with some well-advertised introgressions such as Denisovan genes across East Asia, archaic genes into San or non-tree mixing between Oceanians, East Asians and Europeans. The general result is that these less appreciated and surprising archaic splits have just as much or more support in genome sequence data. Further, evaluation confirms the hypothesis that archaic genes are much rarer on modern X chromosomes, and may even be near totally absent, suggesting strong selection against their introgression. Modeling of relative split weights allows an inference of the proportion of the genome the Denisovan seems to have gotten from an older archaic, and the best estimate is around 2%. Using a mix of quantitative and qualitative morphological data and novel phylogenetic methods, robust support is found for multiple distinct middle Pleistocene lineages. Of these, fossil hominids such as SH5, Petralona, and Dali, in particular, look like prime candidates for contributing pre-Neanderthal/Modern archaic genes to Denisovans, while the Jinniu-Shan fossil looks like the best candidate for a close relative of the Denisovan. That the Papuans might have received some truly archaic genes appears a good possibility and they might even be from *Homo erectus*.

## 1 Introduction

Christmas 2010 was marked with the announcement that a nearly complete genome of an archaic member of the genus *Homo*, nearly equally distantly related to modern humans (*Homo sapiens*) and Neanderthals (*Homo neanderthalensis*) was sequenced and analyzed (Reich et al. 2010). This data and following analyses such as Skoglund and Jackobsson (2011), Waddell et al. (2011, 2012) and Abi-Rached et al. (2011) have begun to reshape our understanding of the genetics of modern humans.

In this article I reevaluate and test some of the more detailed proposals of gene flow between major lineages of the genus *Homo*. Examples of this are complex autosomal gene flow between Oceania (e.g., Papuans), Europe (e.g., French) and the Far East (e.g., Chinese) lineages (Rasmussen et al., 2011), Neanderthal genes into out-of-Africa modern people (Green et al. 2011), gene flow from the Denisovan lineage to Papuans (Reich et al. 2010), gene flow from an earlier archaic ("erectus-like" for want of a better term) into the Denisovan lineage (Waddell et al. 2011), gene flow from Densiova across Far Eastern Asia (Skoglund and Jackobsson 2011), and gene flow from archaics into San Bushmen (Hammer et al. 2011).

I am particularly interested in further evaluating the evidence for an autosomal transfer from erectus-like into Denisovan as this was strongly supported in the analyses of Waddell et al. (2011). This hypothesis did not show up so much in the P2 statistics used there (which are very similar to the D statistics used by Reich et al. 2011), but was evident in the planar graph analysis and as a deficit of Denisova alleles shared with modern humans in the detailed ML analyses of Waddell et al. (2011, 2012). Being a deficit against the whole background of the data, it suggests that this was not simply an imbalance due to Neanderthal genes moving into Africa, for example (of which there was no evidence either).

The planar graph analyses of Waddell et al. (2011) also highlighted a strange arrangement. The lack of a Denisova+Papuan signal, and in its place a Denisova+Neanderthal+ Papuan signal and an even stronger Chimp+Denisova+Neanderthal+Papuan signal. While the former might simply be the expectation that it was not exactly a Denisovan that mated with Papuan ancestors, but rather a lineage closer to them than Neanderthal, the latter is unexpected. It suggests that Papuans may have picked up archaic genes from an archaic member of *Homo* even older than the Neanderthal/Denisovan/Modern split. For want of a better term we call this too "erectus-like." There were indeed erectus-like hominids in SE Asia for a long time, and the last of these may have been the Hobbits® of Flores (*Homo floresiensis*, Morwood et al. 2004, 2005). Whether the more classically erectus forms of the Ngandong specimens persisted until modern humans entered the area is unclear, since there are no other archaic skulls from the area and time period and the last of the Ngandong specimens may be much older than 200kya (Indriati et al. 2011).

The data in Meyer et al. (2012) looked further at the Denisovan and a wider range of modern humans, but did not process comparable data to Reich et al. (2010) for Neanderthal or Chimp. It does however provide data for some useful comparisons with the results based on the earlier data. In particular, Waddell et al. (2011) called attention to a deficit of archaic alleles on the X chromosome; this can be either negative selection and/or male biased gene flow. This was dubbed the archaic Ron Jeremy/lecherous archaic man Hypothesis, for want of a catchy title. Here it is revisited to determine how well it still holds that there is a clear deficient of archaic alleles on modern X and if selection or biased gene flow seems the more probable cause.

Finally, relying only on morphological data can be a major challenge, and in the past phylogenetic trees of major groups have been radically remodeled in the light of molecular evidence (e.g. the mammalian superorders, Stanhope et al. 1998, Waddell et al. 1999, 2001). Analyzing morphological forms within a species-complex, such as the genus *Homo*, presents its own additional set of difficult challenges. Here I explore quantitative methods of phylogenetic analysis of hominid morphological data. So far, there seem to be no or few phylogenetic analyses of morphological data in the genus *Homo*, other than parsimony and the bootstrap, that allow indication of the robustness of clades. Further, parsimony is not easily applicable to the very detailed 3D measurements now possible and becoming available. Applications of novel methods of analyzing distance data may offer a way forward. The data set used is an extensive collection of 3D landmarks and character states from Mounier et al. (2011), and this reanalysis using methods from Waddell and Azad (2009), and Waddell et al. (2010) offers some fairly well supported conclusions regarding the critical middle Pleistocene period of human evolution.

Please note, herein, eight main lineages are assessed, and these are Chimp(C), Densiovan(D), Neanderthal(N), San(S), Yoruba(Y), French(F), Han(H) and Papuan(P). Thus, a split or edge referenced as DNP, is Denisovan+Neanderthal+Papuan versus which ever other taxa are in the analysis. Chimp is the outgroup, so some splits can be polarized that way. In other situations, general biology allows a suggestion of polarity or the direction of gene flow, while in other situations, an over expression of a pattern such as NF, in relation to the full model site pattern spectrum can indicate polarity.

## 2 Materials and Methods

The data used here come from a number of sources. Genomic sequences of eight genomes, including a Neanderthal and Denisovan, are from Reich et al. (2010) as communicated and used in Waddell et al. (2011). Distances based upon the sequencing and resequencing of 12 hominids from tables S18 and S19 were selected from Meyer et al. (2012). These were kindly resupplied by Martin Kircher to full precision. Finally, morphological data (both quantitative and qualitative) are from Mounier et al. (2011) with 3D landmark data kindly communicated by Aure´lien Mounier. These were processed to pairwise Procrustes distances (3D landmarks) and an L1 or Hamming metric (discrete character states) with assistance from Yunsung Kim.

The distance data from tables S18 and S19 of Meyer et al. (2012) were also used. It was confirmed that in that data, transition and transversion distances were performing very much the same, and they were added together to give what is effectively a Hamming, or p-distance (which is effectively the same as a pi distance under the infinite sites model). Martin Kircher kindly provided this same table of distances to full precision (a further 3 decimal places), but this made a very small difference to the results of analyses that were run on both versions of the data.

Morphological data on a wide range of fossil skulls of the genus *Homo* were obtained from Mounier et al. (2011), with 3D landmark measurements kindly provided by Aure´lien Mounier. These were then used to produce a matrix of pairwise “Procrustes” or minimum Euclidean distances (after centering, scaling and rotation), so that they register differences in shape irrespective of size. The discrete character data of the same skulls was processed to produce a mismatch or L1 distance. These were scaled to the same mean then added together to produce the distance matrix of fossil skulls used for analysis in figures 9 and 10.

Tree inference used PAUP* (Swofford 2000, versions b108 to a128) and all fitting of weighted least squares methods (Swofford et al. 1996) involved the logical constraint that the edge lengths must be non-negative (the WLS+ condition). Planar graphs were inferred using the NeighborNet algorithm to select a series of splits conforming to a circular splits system (Bryant and Moulton), followed by reassessment of the edge (split) weights using WLS+. At this stage a split can disappear from the graph if its weight goes to zero. This was done using SplitsTree4 (Huson and Bryant 2006). To perform residual resampling and record associated statistics scripts developed in Waddell and Azad (2009), Waddell et al. (2010) and Waddell et al. (2011) were used. Spreadsheet calculations and numerical optimizations were performed in Microsoft Excel.

## 3 Results

### 3.1 Comparing signals in the 2010 versus 2012 Denisova Data

The first task is to assess the 2010 Reich et al. data used in Waddell et al. (2011) in relation to the refined, but not strictly comparable data of Meyer et al. (2012). The 2010 data used were complete site patterns that passed filter as communicated by Nick Patterson (see Table A1). The 2012 data used are the table of pairwise distances reported in table S14 and S16 of Meyer et al. (2012), with extra digits provided by Martin Kircher, followed by summing the transitions and transversions together. For a valid comparison, each distance data set was whittled down to just the common genomes (DSYFHP).

For each of these four distance matrices, a WLS+ P=1 tree was fitted, followed by fitting a WLS+ P=1 planar graph, with splits selected by the Neighbor Net algorithm (Table 1). It is the ig%SD (Waddell et al. 2010) fit statistic that is of particular interest, since it measures the quality of fit to the internal edges or informative splits. This statistic is much less prone to the impact of sequencing errors, which affect a large fraction of the singleton site patterns in Reich et al. (2010) and may still be having a detectable effect with the data from Meyer et al. (2012). The overall result is that while the fit of a tree to all four data sets is not great, three of the data sets fit the planar model very well. These are the autosomal 2010 and 2012 distances and the 2012 X-chromosome distances. For these three data sets, fit criteria such as AIC, AICc and BIC clearly favor the planar model. Further, the 2010 autosomal data is nearly as good as the 2012 data set at discerning internal structure (ig%SD 4.34% versus 3.86%) suggesting that despite many more sites being included (~50% rather than ~10% of the genome) and much deeper sequence coverage, both data sets are roughly as good as each other at determining informative splits. Indeed, in terms of the BIC statistic, the 2010 autosomal model seems to be a better model than 2012 model. This is helpful, as we are particular interested in extending the results on archaic interbreeding and the data of Meyer et al. (2012) did not have comparable data for chimp or Neanderthal, which are highly informative.

Table 1. The fit of autosomal distances and X chromosome distances for just the six taxa (DSYFHP) common to the 2010 and 2012 data sets. Fit is compared on a tree with WLS+ $P$=1 and a NeighborNet selected WLS+ $P$=1 planar graph. The heading T2010auto indicates it is the fit of a tree to the 2010 autosomal distance data (which was based on transversions only), while 2012X indicates it is a planar graph (lack of T) for the 2012X derived distance data from Meyer et al. 2012 (as reported with extra digits of precision). The reported values are: GeoMean; geometric mean of the observed distances, AriMean; corresponding arithmetic mean, SS; residual sum of squares, MSE; SS/number of distances, g%SD; geometric %SD misfit, %SD, arithmetic percentage misfit, ig%SD; g%SD divided by the proportion of the graph comprised of internal edges, i%SD; the same, but for %SD, lnL; log likelihood, AIC; Akaike Information Criterion, AICc; small sample adjusted AIC, BIC; Bayesian Information Criterion.

| | T2010auto | T2012auto | T2010X | T2012X | 2010auto | 2012auto | 2010X | 2012X |
|---|---|---|---|---|---|---|---|---|
| GeoMean | 715.59 | 1229.88 | 683.15 | 864.39 | 715.59 | 1229.88 | 683.15 | 864.39 |
| AriMean | 716.27 | 1249.99 | 683.73 | 894.95 | 716.27 | 1249.99 | 683.73 | 894.95 |
| SS | 0.13 | 0.66 | 0.09 | 0.25 | 0.00 | 0.01 | 0.02 | 0.01 |
| MSE | 0.01 | 0.04 | 0.01 | 0.02 | 0.00 | 0.00 | 0.00 | 0.00 |
| g%SD | 0.55 | 0.95 | 0.47 | 0.70 | 0.11 | 0.20 | 0.27 | 0.26 |
| %SD | 0.55 | 0.94 | 0.47 | 0.69 | 0.11 | 0.20 | 0.27 | 0.26 |
| ig%SD | 27.40 | 21.76 | 20.22 | 9.36 | 4.34 | 3.86 | 10.21 | 3.24 |
| i%SD | 27.38 | 21.58 | 20.22 | 9.19 | 4.34 | 3.83 | 10.20 | 3.19 |
| lnL | -35.05 | -51.26 | -31.89 | -41.39 | -2.40 | -14.70 | -18.65 | -13.30 |
| AIC | 88.10 | 120.51 | 81.78 | 100.78 | 30.81 | 57.39 | 61.29 | 54.60 |
| AICc | 124.10 | 156.51 | 117.78 | 136.78 | 394.81 | nan | 217.29 | nan |
| BIC | 94.47 | 126.89 | 88.15 | 107.15 | 40.01 | 67.30 | 69.79 | 64.51 |

The situation with the X chromosome is somewhat different. The 2012 data fits markedly better than all the other data sets to a tree, and about as well at the autosomal data sets to a planar graph. In contrast the 2010 X chromosome data fits a planar graph somewhat poorly. This suggests that the 2012 X chromosome distances in particular are worth examining to test the predictions of Waddell et al. (2011) that informative site patterns linking archaics to moderns are particularly rare in the case of the Denisova/Papuan link due to either male biased gene flow and/or selection. There is already a clear hint from table 1 that the 2012 X chromosome data is more tree-like than the autosomal data, which itself hints at less horizontal gene transfer. Note, using just the transversion distances for the 2012 data the results are very similar.

The actual models recovered are shown in figure 1, with support values using residual resampling based on adding errors of the form $N(0, \sigma^2)$ to the expected or model distances. In terms of the trees recovered, the tree itself was always the same with support values of 100 on all edges, except the short Han+Papuan edge (results not shown).

The planar graphs for each set of data are shown in figure1. The 2010 data shows a result that is very similar to that of Waddell et al. (2011), as would be hoped. The Denisova+Papuan split is strongly supported as is the non-tree FH signal. The 2012 autosomal data yields a very similar planar graph, with similar support values, except that the 2012 data resolves at moderately high support a couple of additional splits. One of these is a DHP/SYF split that is consistent with Han appearing to have more archaic alleles than F as seen in analyses of both Reich et al. (2010) and Waddell et al. (2011). Meyer et al. (2012) claim this as significant, but the support value here is more equivocal. A YF and YFH splits also appear with moderate support but low weight.

The 2010X and 2012X data yield graphs that are somewhat different to each other and the autosomes. While the 2010X chromosome graph looks somewhat like the autosomal graphs, with a weaker DP but stronger FH signal, the 2012X data adopts a different configuration. The DP signal is completely gone, as is the FH signal, and in their place are a myriad of smaller signals. In one sense the 2012 data confirms the findings of Waddell et al. (2011) that chr X is deficient in DP archaic alleles in particular, and in another sense suggests that the results would have been even stronger but for issues with the data. One possibility is that in the 2010 data processing, short reads from autosomes were incorrectly assigned to the X chromosome much more often than in the 2012 data processing.

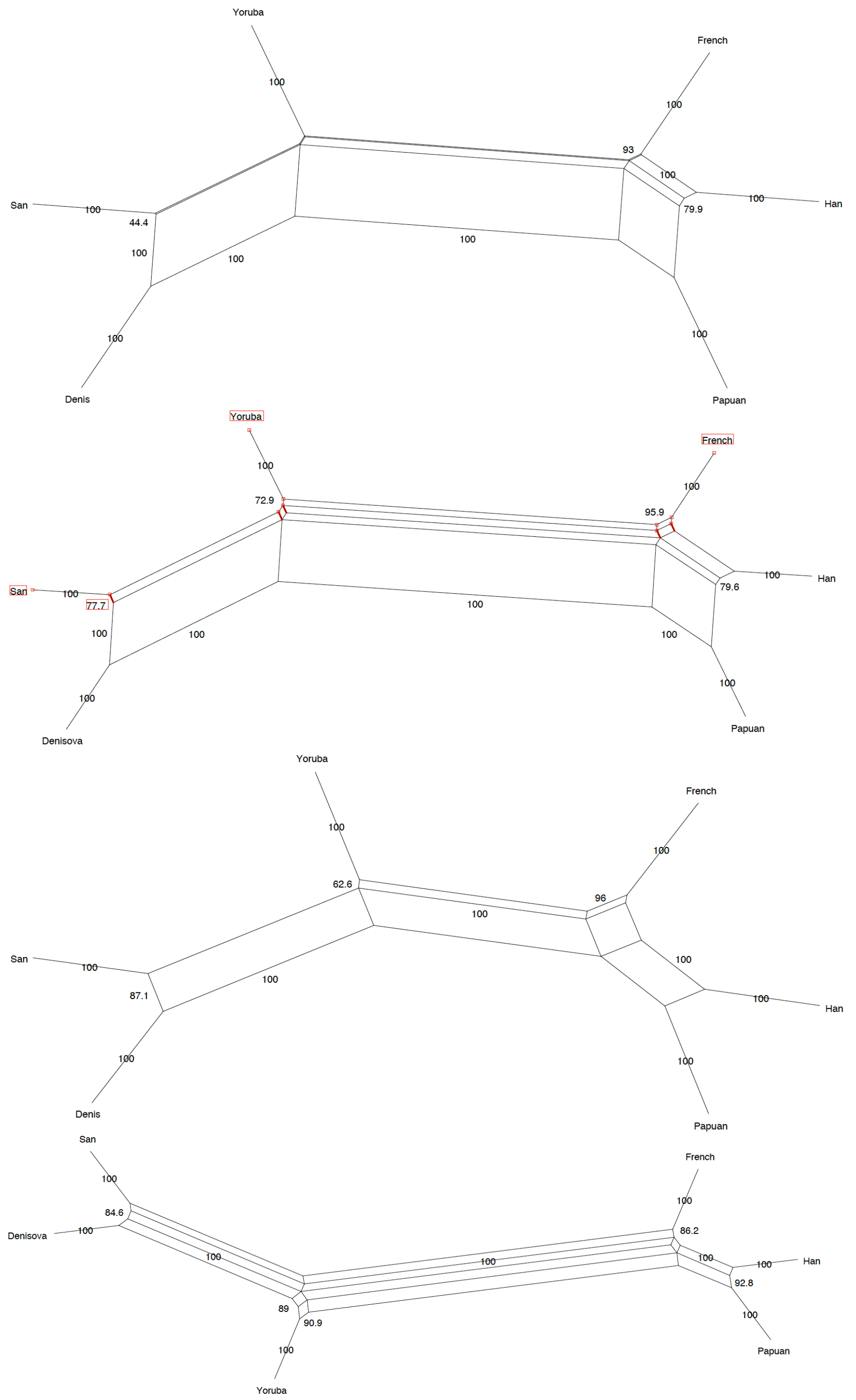

Figure 1. WLS+ $P$=1 (with NeighborNet) Residual Resampling results of the subset of distance data common to Reich et al. (2010) and Meyer et al. (2012). (a). 2010 autosome distances. (b) 2012 autosomes. (c) 2010 X chromosome. (d) 2012 X-chromosome. The relative fit values are ig%SD 4.34, 3.86, 10.21, and 3.24, respectively.

**3.2 Planar model residuals of the 2010 autosomal data**

In order to being to look in greater detail at archaic interbreeding in the autosomal data and further test hypotheses such as Reich et al. (2010) of Denisovans interbreeding with Papuans and Waddell et al. (2011) of Denisovans interbreeding with an even earlier archaic, it is useful to pick apart the residuals to the model, as done in Waddell and Azad (2009).

Table 2 shows the signed squared weighted residuals for the WLS+ P=1 model fitted to the data. The weight being used, P=1, approximates the log likelihood of the multinomial model based upon pairwise distances. The impact of a squared residuals upon overall fit is proportional to the inverse of the distance that they apply to. In this data, all pairwise distances are very roughly the same except the long distance out to the chimp, which is approximately four times as large as the others. Underlined in the table are the two largest contributors to misfit. The largest is negative and it indicates that Papuan would like to be closer to the Denisovan. Since the raw residual is effectively a negative distance, this outcome can come about by the addition of uniquely derived alleles to just Papuan and Denisovan.

The large positive residual for the Neanderthal/Papuan distance is almost the same size as the negative residual of the Denisovan to Papuan. Taken together, and noting that the splits in a tree or a planar diagram are equivalent to orthogonal axes in an L1 (or Manhattan) distance space, then the end result seems to be the desire for a swap, which is interpreted as more Denisova-like derived alleles into Paupuan in exchange for less Neanderthal-like derived alleles, with respect to what the planar diagram can achieve. The table also shows the absolute sum of weighted residuals and in this case, Denisovan then Papuan are the worst fit. Summing the squared weighted residuals measures deviation in terms of the WLS+ fit criterion, and in this case Denisovan remains the worst fitting, with Papuan close behind.

Figure 2 shows the results in graphical form. Figure 2 shows the residuals to the WLS+ P=1 planar model in raw form, then as weighted residuals adjusted by the standard deviation and, finally, as the squared weighted residuals adjusted by the inverse of the variance (which is the total sum of weighted squared residuals divided by the remaining degrees of freedom). The inverted pattern of Denisovan wanting to be closer to Papuan and conversely further from Han and French, versus the Neanderthal wanting to be further from Papuan but closer to Han and French suggests that the archaic versus Han and French residuals seen here may be largely a side-effect of a best fit in the presence of the much larger unaccounted for residuals of archaics to Papuan. Also notable in the graphs is that Neanderthal and Denisovan would prefer to be slightly closer to each other, but the exact cause of this is unclear and it too might be the result of compromises due to the largest misfitting residuals.

What is also notable is that where chimp fits into the picture is not complicated by large residuals. Indeed, in terms of squared weighted residuals, chimp is very well located, and even the raw residuals do not show strong discrepancies. Why that becomes so interesting is that the dominant split of Papuans with archaics is not Neanderthal nor Denisova alone, nor together, but both plus Chimp (Waddell et al. 2011), suggesting the possibility that whatever archaic interbreed with Papuan ancestors had more input from earlier archaics than even the Denisovan individual.

Table 2. The signed squared weighted residuals after fitting a WLS+ P=1 planar model to the 2010 autosomal data. The final column is the row sum of the absolute values, and an indicator of which taxa are fitting worst. The two largest squared residuals are underlined and the largest row sums are in bold.

| | Chimp | Denis | Neander | San | Yoruba | French | Han | Papuan | Abs Sum |
|---|---|---|---|---|---|---|---|---|---|
| Chimp | 0.0 | 0.0 | 0.0 | 0.0 | 0.0 | 0.0 | 0.0 | 0.0 | 0.17 |
| Denis | 0.0 | 0.0 | -0.4 | 0.0 | 0.0 | 0.4 | 0.6 | -1.3 | **2.73** |
| Neander | 0.0 | -0.4 | 0.0 | 0.0 | 0.0 | -0.1 | -0.4 | 1.1 | 2.07 |
| San | 0.0 | 0.0 | 0.0 | 0.0 | 0.0 | 0.0 | 0.0 | 0.0 | 0.14 |
| Yoruba | 0.0 | 0.0 | 0.0 | 0.0 | 0.0 | 0.0 | 0.0 | 0.0 | 0.15 |
| French | 0.0 | 0.4 | -0.1 | 0.0 | 0.0 | 0.0 | -0.1 | 0.0 | 0.80 |
| Han | 0.0 | 0.6 | -0.4 | 0.0 | 0.0 | -0.1 | 0.0 | -0.1 | 1.41 |
| Papuan | 0.0 | -1.3 | 1.1 | 0.0 | 0.0 | 0.0 | -0.1 | 0.0 | **2.54** |

Figure 2. WLS+ $P$=1 2010 autosomal residuals shown as a plot. (a) The raw residuals, which, barring convergences or parallelisms, can be considered proportional to derived alleles. (b) The weighted residuals (adjusted by the inverse of the estimated standard deviation of that distance). (c) The squared weighted residuals, which are proportional to the likelihood function being used. The two largest residuals indicate that Papuan wants to be further from Neanderthal than the planar diagram allows, and about this much closer to the Denisovan.

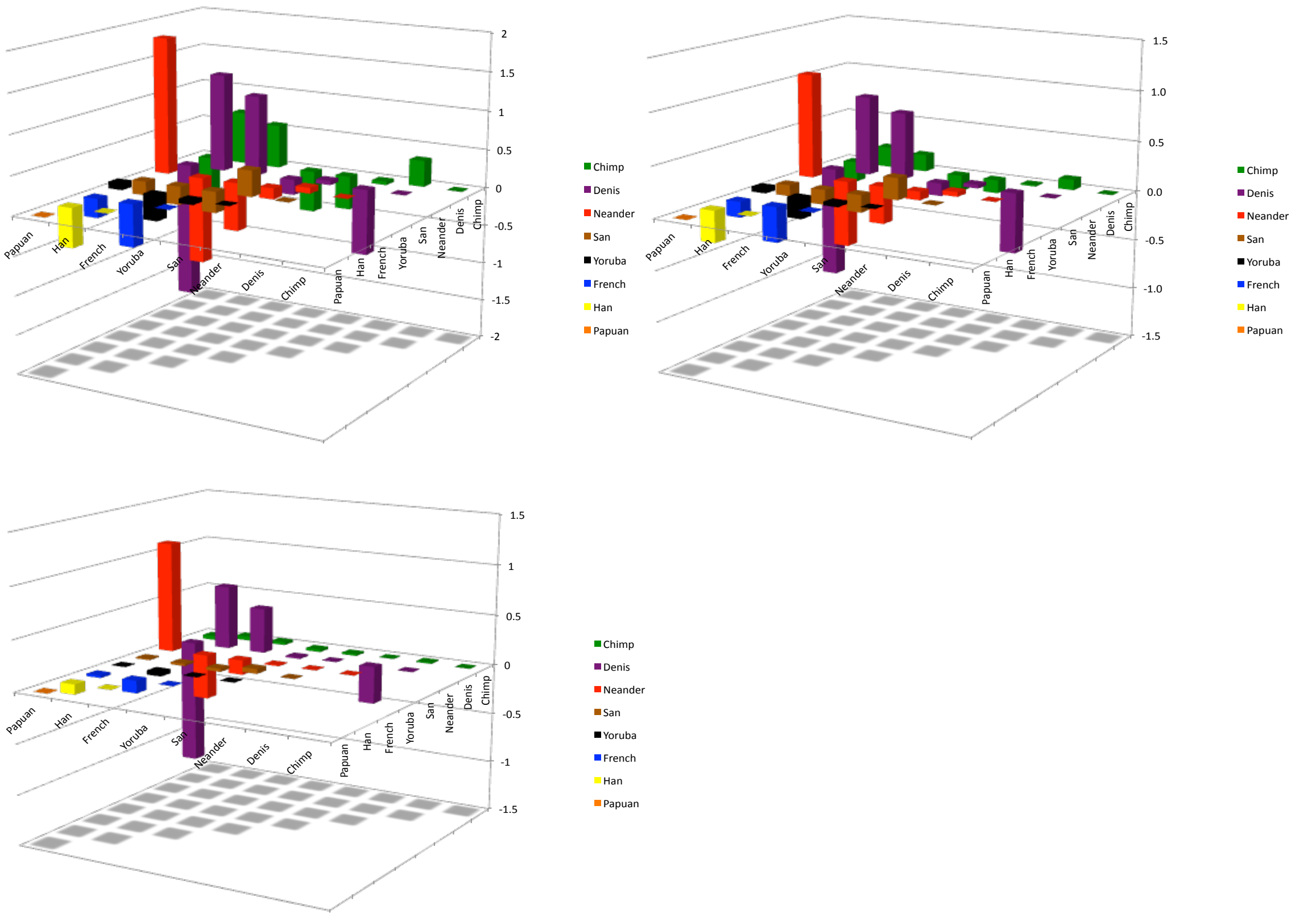

## 3.3 No X please! We are modern humans.

Before delving more deeply into the exact nature of archaic interbreeding suggested by the 2010 data as analyzed in Waddell (2011) it is important to address what the X chromosome is doing. For this purpose it is helpful to use the 12 full modern genomes sequenced by Meyer et al. (2012). There, in addition to D, S, Y, F, H and P, there is a Mbuti Pygmy, a Mandenka West African, a Dinka North East African, a Sardinian (European), a Karitiana (Native South American) and a Dai (Southern Chinese) genome to give a richer picture.

Results of fitting chromosome X distances to planar diagrams are shown in figure 3. The results for transition, transversion and the total nucleotide distance are highly consistent amongst the out of Africa individuals. They suggest that, in addition to the expected tree structure, X chromosomal material was horizontally transferred between the ancestral Karitiana and Han populations, and possibly between the ancestors of Sardinian and Karitiana. Within Africa, there is a weak split of Denisovana + Mbuti, which could be a small amount of archaic X into the Mbuti X, but more likely, San lineage females moving out of Southern Africa (there is some good prior genetic evidence of Mbuti being mixed). The ordering of the divergences of Yoruba, Mandenka and Dinka are poorly resolved.

Most notably, there is no evidence for a signal linking Denisova to Papuan. Indeed, even with heavily reduced taxa sets such as Dai, Papuan, Denisova, San, there is no evidence for this signal. The same seems to hold for other combinations, randomly examined, such as Denisovan,

San, Mandenka, Dai and French, with no additional split of Denisova with any of these, but a small non-tree split linking Dai and Mandenka (which might indicate a effective threshold of resolution). This corroborates the hypothesis of a reduced frequency of Denisova X derived alleles in the Papuan individual (Waddell et al. 2011).

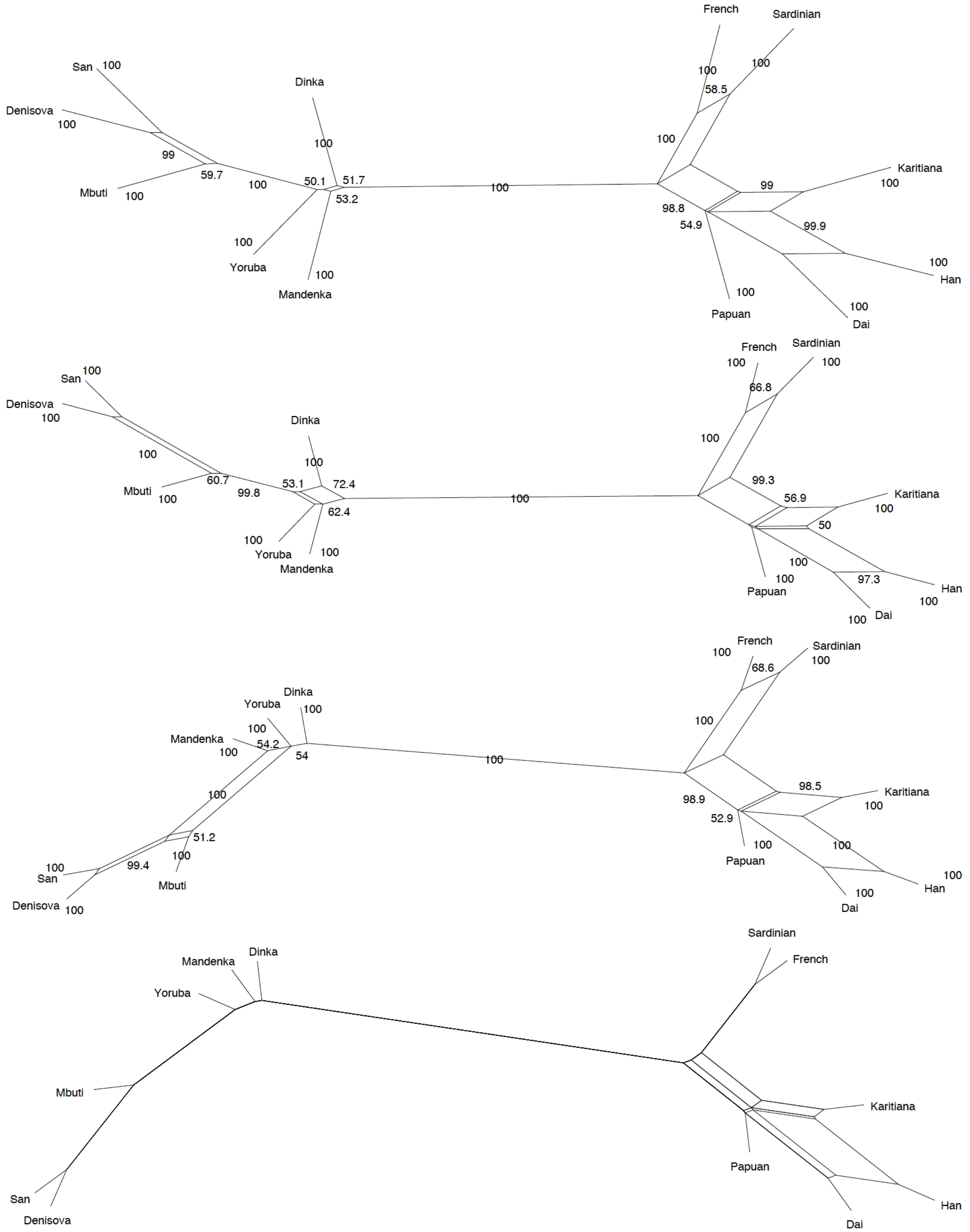

Figure 3. WLS+ *P*=1 Planar Diagrams with a Residual Resampling based threshold of 0.5 based on X chromosome-based distances. (a) Transversion distances. (b) Transition distances. (c) Combined distances. (d) Combined X distances fitted onto the Autosomal NeighborNet graph (note, most of the interesting and well supported non-tree splits in the autosomal graph shrink to zero with X).

In terms of fit, the planar diagrams are judged a better description of the X chromosome distance for AIC and BIC for transversions, and only AIC for transitions, and by none of the criteria for the combined distances. That is, a single "species" tree is frequently assessed a better model than a NeighborNet selected planar graph for this data, which is further evidence of a lack of archaic genes moving across to moderns. This is in strong contrast to the same analyses of the autosomal data, as seen earlier.

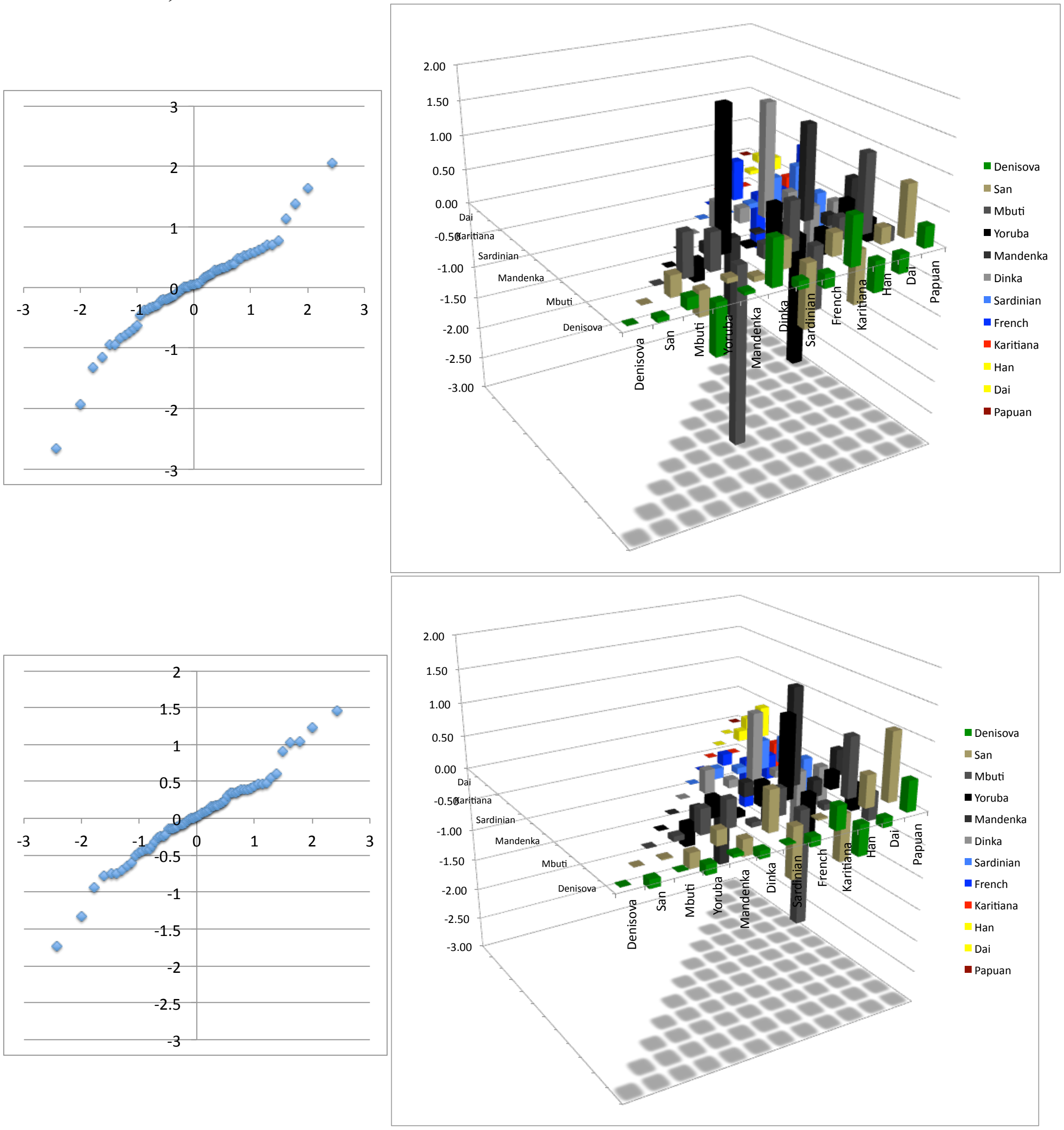

Figure 4. (a) Top left, a Q-Q plot of the standardized residuals for the NeighborNet WLS+ $P$=1 model for chromosome X distances versus their expected values. (b). 3D plot of the standardized residuals. The negative outliers are Mbuti and Dinka (-2.66) and Yoruba and Karitiana (-1.92). These are associated with the large positive deviations Dinka and Yoruba (2.05) and Karitiana and Dinka (1.64). Note, there is definitely no negative Denisova+Papuan residual, and only a weak non-significant looking negative Denisova+Han residual. In strong contrast to the autosomes, there is no evidence of archaic gene flow into moderns on the X chromosome, with all such residuals falling well within the pack of the Q-Q plot. (c) The Q-Q plot after adding back in twice the two largest negative residuals to just those distances (approximately fitting a Mbuti+ Dinka and Yoruba+Karitiana split) then reestimating the network. Importantly, there are no other such obvious outliers and no evidence of strong negative splits with the archaic become evident.

It is also possible to consider how genes on X fit a network of 44 splits obtained with NeighborNet applied to autosomal distances (with a large outlying residual of Denisova to Mandenka removed, but with a split of Denisovan+Papuan pre-selected, network not shown). The results are that most of the non-tree splits have an estimated length of zero, as seen in figure 3d. Informative tree splits in the graph have estimated edge lengths that are similar to those obtained with a tree fitted to X (not shown). For example, the split of Denisova+San from all others has length 27.21 compared to 27.60. Other tree splits compared to their tree values are Denisova+San+Mbuti 32.02 (34.66), Dinka+modern-non-Africans 1.64 (3.47), Modern non-Africans 107.69 (109.80), Sardinian+French 21.90 (25.43), Papuan plus East Asians 19.45 (18.16), and Han+Dai 26.63 (19.33). The only major non-tree autosomal splits that remain in evidence are Karitiana +Han with weight 16.02 and Sardinian+French+Karitiana with weight 3.14. Thus, there is no apparent hidden evidence at all for the Denisova+Papuan split, which here (Figure 3d) is reconstructed with length zero.

A detailed examination of the residuals of the X data to its own preferred WLS+ P=1 Planar diagram is useful to further address the issue (figure 4a and b). This first model suggests unmodeled interbreeding of Mbuti and Dinka plus Yoruba and Karitiana. If we fit the first pair by adding back to the original distance matrix –twice the residual, the lnL improves from -219.57 to 202.35. The number of splits in the NeighborNet increases to 35, and k = 36. The AIC and BIC fit values are favorable to the new model, but the AICc value is not. Adding in a parameter to model Yoruba and Karitiana mating (genes from an African worker/slave going into the Amazon?), the lnL only improves slightly to -200.94. The new fit values are AIC 467.90, AICc 533.90, and BIC 540.16. These are all improvements, and all are better than the simpler model. They are also all better than the tree, but that is about all that is in evidence for in terms of unaccounted for splits and is justified in this data. There is no evidence for Denisovan-like genes introgressing into moderns. The Denisovan genome appears to be a reasonable proxy for a wide range of archaic genes (although not as powerful as having the exact archaic genome represented, if, for example, the introgression was Neanderthal) and there is no evidence of enrichment in far easterners or non-Africans compared to Africans (where the Denisovan appears to be a very appropriate genome to have).

### 3.4 Interrogating autosomal splits via the removal of one taxon at a time

As mentioned previously, planar graphs are an extension of trees, but are not fully general graphs. One way of detecting hidden or missing splits is to remove each taxon in turn and look at the change in fit. The data used are the 2010 transversion only distances.

Again, a particularly informative statistic here is expected to be the ig%SD fit statistic. This had a value of 6.96 in its unbiased form (adjusted for the degrees of freedom) for the full data set with WLS+ P=1. Removing taxa S, Y and F saw this value increase slightly to around 7.5, indicating that these taxa were essentially not contributing to misfit of the planar model (table 3). In contrast, removal of P saw a substantial improvement in fit with ig%SD dropping dramatically to 2.71 (table 3). The probability of this much improvement by chance, can be assessed approximately, by noting that the mean and SE of the eight tip residual resampling replicates were, respectively, 7.35 (up from the original fit of 6.96 due to uncorrected biases) and 2.28 respectively. Thus a fit of 2.71 would seem to be just outside the approximate 95% confidence interval of +/- 2 SE, or 2.79-11.91). A similar result is obtained if the residual resampling replicates for S, Y or F is used as the baseline (the three seven taxon sets with minimal change from the original data). (Note, the larger than expected SE for replicates with N removed is due to a few replicates fitting very poorly, perhaps due to the reconstruction method imposing a non-negative edge length boundary). Since the fit improves most with Papuan removed, this suggests the planar diagram is having most trouble fully capturing its genomic relationships. This is consistent with what was seen earlier examining the residuals.

Removal of Denisova, Neanderthal and Chimp singly, also results in substantial improvements in fit. Why their removal improves the fit is unclear, but it could be removing the effect of archaics, which may be interacting with Papuan in complex ways that the planar diagram cannot explain all at once. However, the focus is on Papuan, since it seems to show a complex pattern of splits with multiple archaics, including a Chimp, Neanderthal, Denisovan + Papuan split (which suggests one or more lineages of an as yet unsampled and particularly archaic *Homo* lineage).

Table 3. The fit of autosomal distances to a NeighborNet WLS+ $P$=1 planar graph with each taxon removed in turn. For terms used, except SE, see the caption of table 1. For ig%SD removal of Papuan produces the best fit, followed by Denisovan, Neanderthal and Chimp. SE is the standard error of the ig%SD statistic over 1000 residual resampling estimates.

| taxon | C | D | N | S | Y | F | H | P |
|---|---|---|---|---|---|---|---|---|
| GeoMean | 725.92 | 1132.76 | 1093.71 | 1112.40 | 1102.65 | 1106.32 | 1093.17 | 1091.84 |
| AriMean | 728.44 | 1444.42 | 1414.30 | 1428.31 | 1420.46 | 1422.50 | 1412.24 | 1411.40 |
| SS | 0.0171 | 0.0019 | 0.0025 | 0.0148 | 0.0147 | 0.0144 | 0.0112 | 0.0024 |
| MSE | 0.0008 | 0.0001 | 0.0001 | 0.0007 | 0.0007 | 0.0007 | 0.0005 | 0.0001 |
| g%SD | 0.22 | 0.06 | 0.08 | 0.21 | 0.21 | 0.21 | 0.16 | 0.07 |
| %SD | 0.22 | 0.06 | 0.07 | 0.19 | 0.19 | 0.18 | 0.14 | 0.07 |
| ig%SD | 3.99 | 3.40 | 3.62 | 7.49 | 7.45 | 7.49 | 5.94 | **2.71** |
| SE | 1.53 | 1.55 | 2.31 | 2.80 | 2.79 | 2.84 | 2.59 | 1.49 |
| i%SD | 3.98 | 3.01 | 3.19 | 6.61 | 6.56 | 6.61 | 5.22 | 2.39 |
| lnL | -24.30 | -5.90 | -8.23 | -27.21 | 2.70 | -26.91 | -24.10 | -8.10 |
| AIC | 80.59 | 45.80 | 50.46 | 90.42 | 90.19 | 89.82 | 82.19 | 50.20 |
| AICc | 216.59 | 249.80 | 254.46 | 432.42 | 432.19 | 431.82 | 286.19 | 254.20 |
| BIC | 97.31 | 63.55 | 68.21 | 109.23 | 109.00 | 108.62 | 99.95 | 67.96 |

To examine more closely what is going on, it is possible to not only look at the individual planar diagrams with residual resampling support values, but also to plot out the support (its weight) for splits as each taxa is removed. We first look at the individual diagrams.

Figure 5 shows the eight possible seven-taxon planar diagrams created by removing one of the original eight genomes, ordered by their goodness of fit according to the ig%SD value. They are not exactly at the same scale, but they are oriented similarly. All terminal edges have been reduced to a small, constant value so the internal structure is clear. With Papuan removed, the ig%SD fit is most improved and this in turn reduces the overall error which residual resampling replicates. The support for a Chimp+Denisovan split is now over 99%, suggesting that Papuan's complex genetic history outside of the modern human ancestor may have been competing with this signal and obscuring it. The split of Nenderthal with the out-of-Africa people is also well supported with 95% support. Other non-tree splits remain equivocal, but there are two faint signals suggesting that Han is enriched in derived alleles like those of Neanderthal and like those of a mixture of Neanderthal and Denisova, both with ~45% residual resampling support. The former agrees with the findings in Reich et al. (2010), Waddell et al. (2011) and Meyer et al. (2012) (the last most conclusively) of Neanderthal enrichment in East Eurasians over West Eurasians, while the latter is similar to the suggestion of Skoglund and Jakobsson (2011) of some Denisovan-like alleles spread wide in the far-East.

With Denisovan removed, the Neanderthal with out-of-Africa people signal increases to 98% while the suggestion of more Neanderthal like alleles in East Eurasia also increases. A Papuan+Chimp+Neanderthal signal is notably the largest non-tree signal with 100% support. A French+Han signal, while not so large is well supported (98%), along with a weaker Yoruba+ French+Han signal (support 79%). A small San+Chimp signal emerges with 62% support. A Chimp+San signal may be consistent with the findings of Hammer et al. (2011) for San archaic interbreeding, if the archaic kept an excess of gene lineages (in relation to modern humans) markedly older than the Neanderthal/modern divergence.

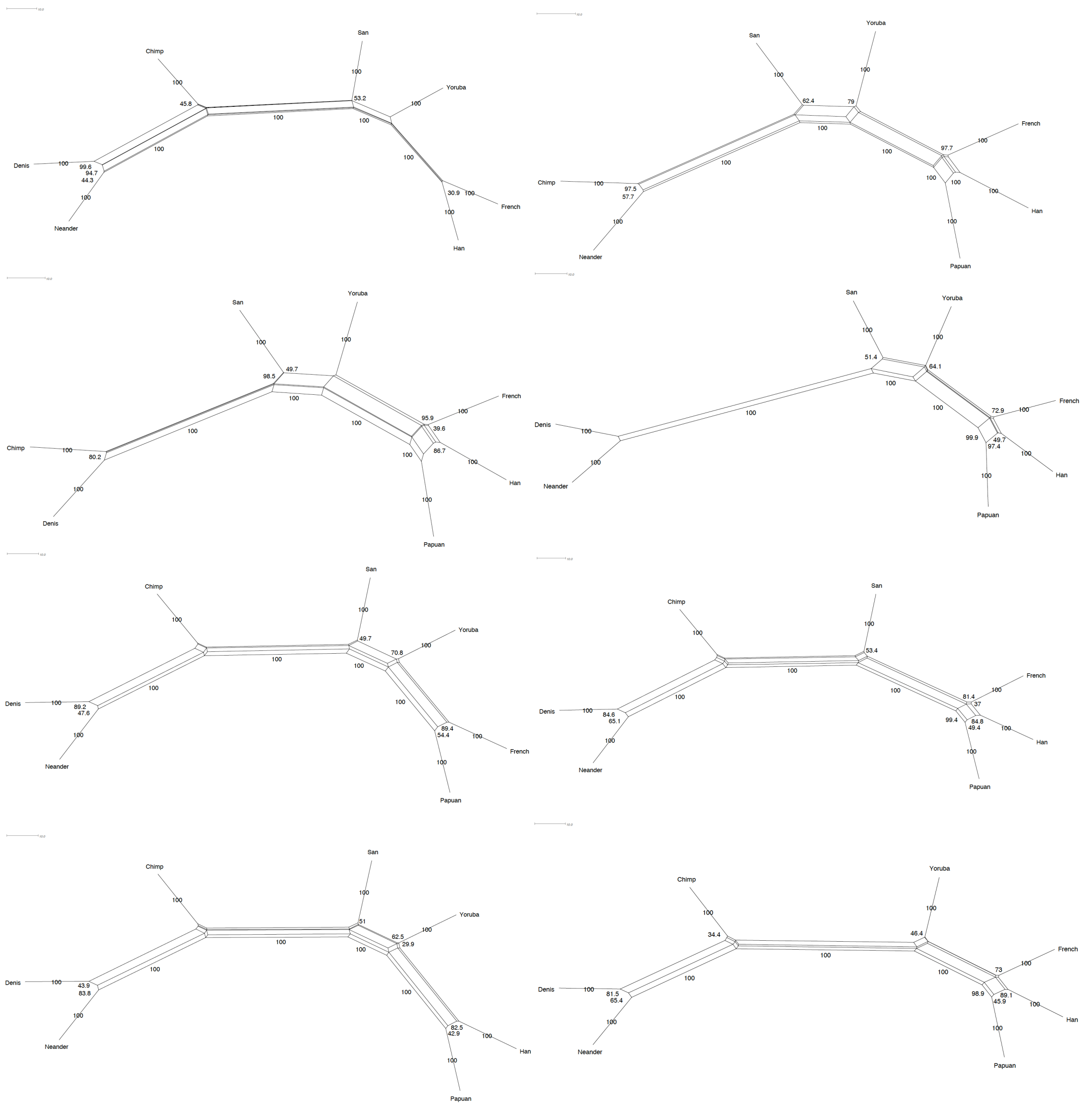

Figure 5. Each WLS+ P=1 planar graphs created by removing one taxon at a time. The order from left to right following down the page is from best to worst fit, that is P, D, N, C, H, Y, F and S removed in that order.

Figure 6 shows the changing weight of splits when different taxa are removed. Note that tree edge weights are very stable except for an increase in FHP when N is removed, an increase in ND when P is removed and increases in all moderns joined together when either P or N is removed.

The non-tree split weights are seen to be more variable in general than the tree signals (Figure 6). However, the CD signal of erectus-like into Denisova is very stable, and hence does not seem to be artifactually affected by the inclusion of other taxa. The enigmatic CDNP signal remains at least as large as CD and peaks when S or N are removed. Removal of N might make the signal more precise in allowing it to represent more of D plus a lineage off the common ancestor of D and N plus an apparently more archaic hominin branching off along the edge to chimp. It is of course possible that CDNP is in fact better described as SYFH. However, given what is known of human origins, particularly the dominant species tree and the long geographic isolation of San in Southern Africa, such a strong SYFH signal seems unlikely. This is corroborated by a much smaller YFH signal than CDNP, plus YF or FH both being relatively small also. All these would seem to be better in line with biogeography and archeological history, than a major out of Africa (but not Oceania) movement of the San genes after Papuans entered Oceania.

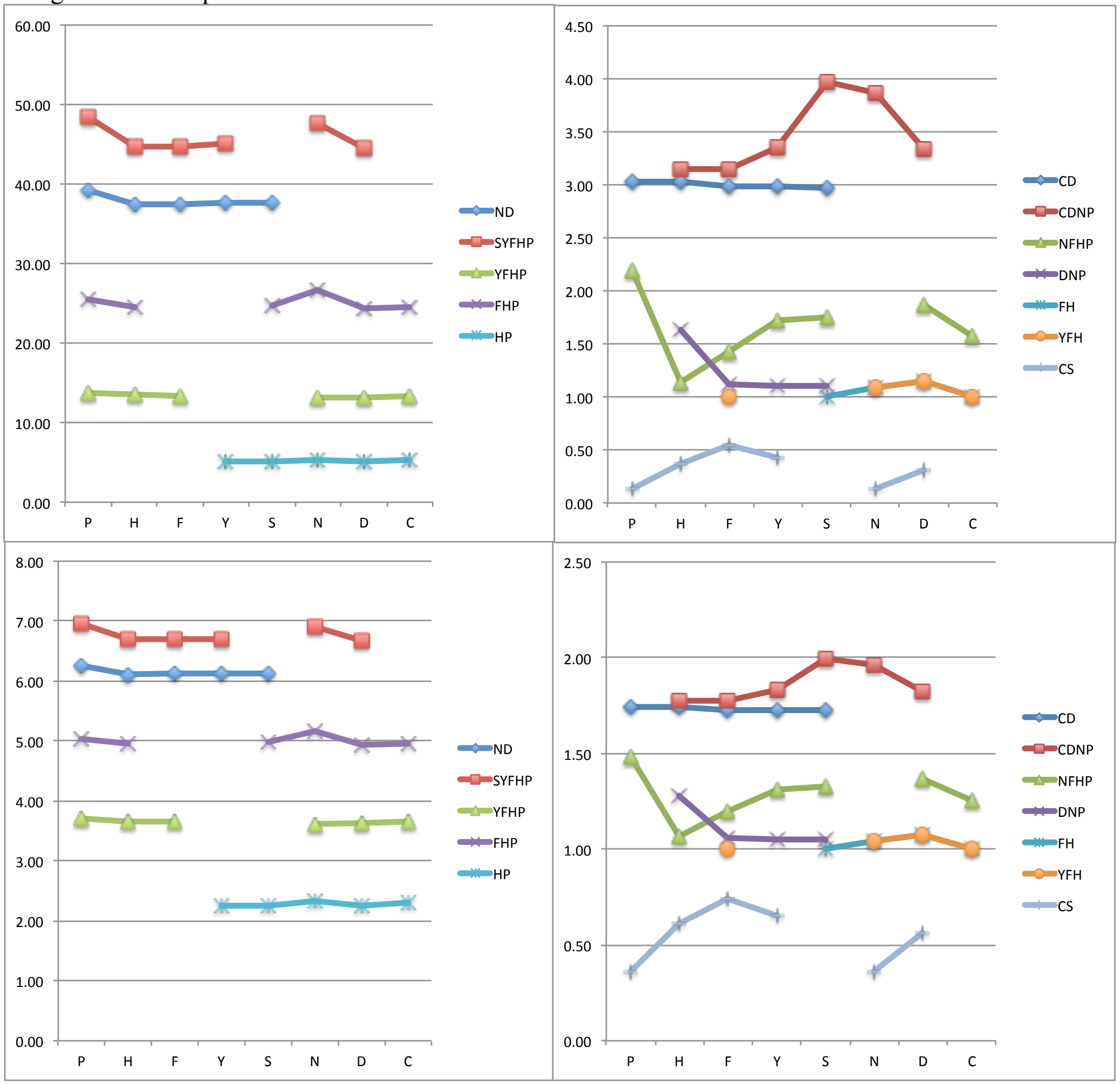

Figure 6. The strength of particular splits for WLS+ P=1 planar diagrams when one taxon of the 8 is left out. (a) Top left, the weights of the main species tree splits. (b) Top right, the weights on non-tree splits. (c) as for (a) but a square root transform has been applied to moderate differences in expected sampling variance. (d) As for (b) but a square root transform has been applied to moderate differences in expected sampling variance. In general, a split is shown if removal of a specific taxon does not directly disrupt its estimation or if removal of a taxon does not make it isomorphic with another stronger split (for example, with the removal of San, YFH becomes isomorphic with the much stronger split CDNP). Also, note, with no H, YFH becomes YF, which is not shown but has a weight of 1.63.

The highly touted archaic inbreeding non-tree signal NFHP (Green et al. 2010, Reich et al. 2010, Skoglund and Jakobsson 2011, Waddell et al. 2011) and the prominently proposed CS signal (Hammer et al. 2011) are both more erratic than CD or CDNP, while CS is also particularly small. The NFHP signal peaks when P is removed, consistent with the residual that suggests the model desires less N into P. In contrast, it drops when H is removed, consistent with results (e.g., Waddell et al. 2011) that H has more derived N alleles than F (by both the P2D2 type tests and full site-pattern counts with reticulation ML modeling). The CS signal peaks when F is removed, yet it is unclear given current knowledge how F would be affecting this signal.

In terms of residual resampling support, the support values for CD on 1000 replicates are -P:0.996, -H:0.892, -F:0.838, -Y:0.846 and -S:0.815 versus -Y:0.814, -S:0.73, -N:0.959, -D:0.977, and -C:0.729 for the FH signal. At the peak, with minimal interference from specific genomes, the CD signal is 0.966 versus 0.977 for FH. The FH signal too has been a highly touted non-tree signal (for example, Rasmussen et al. 2011), yet the CD split pretty well matches it in support.

Support values for CDNP are also very strong -H:0.894, -F:0.825, -Y:0.848, -S:0.891, -N:0.985 and -D:1.00, this is a peak of 1000 when D is removed. In contrast, the signals for both DNP and CS are much weaker and more variable. For DNP the support is -H:0.544, -F:0.429, Y:0.494, -S:0.459, and -D:0, with the last value (removing D) being included to show that this is not an NP signal and that D is part of the signal, but certainly not sufficient. For the better known CS split, the support numbers are -P:0.458, -H:0.497, -F:0.51, -Y:0.534, -N:0.497, and -D:0.624 which are fairly similar except for a bit of a peak for CS when D is removed.

### 3.5 Quantitative assessment of Denisovan and an earlier archaic interbreeding

The results above corroborate the findings of Waddell et al. (2011) that the Denisovan individual indeed seems to have an appreciable fraction of its autosomal genome from a hominid that diverged prior to the Modern/Neanderthal split. As seem in figure 6 above, with the removal of taxa, this split remains fairly stable. Probably the single best estimate of the size of this CD split comes when Papuan is removed from the data set, and the fit statistic ig%SD markedly improves.

The next step is to infer the spectrum of splits expected from genetic distances with one reticulation. This is much simpler than the detailed coalescent calculations involving multi-way integrals undertaken in Waddell et al. (2011) to infer the spectrum of all possible site patterns (sequence pattern "splits"). Assuming a genetic clock makes the process easier still, but is not essential, and can be relaxed. Please note, I am going to use the term "erectus-like" here, to describe an archaic hominin that split prior to the modern versus Neanderthal split, for want of a better term. As we show later, while in 2011 "erectus-like" seemed an adequate description of all fossils older than ~800,000 years from Asia, our analyses below suggest that the morphology of some hominids in Eurasia in this period of time (~600kya to 1mya) might have been more like that of the SH5, Dali or Petralona fossils, which are typically thought of as later middle Pleistocene forms.

The model being used is shown in figure 7. From this model, it is a fairly straightforward to calculate the expected distance split spectrum. The reticulate model produces two "species" trees, one of the form (E,((D,N),M)) and the other, in proportion P(E), of the form ((E,D),(N,M)). On each of these two trees I need to calculate the edge lengths. Without loss of generality, the edge lengths on these two trees are calculated from the mean or expected coalescent times of a pair of taxa across each internal node of the rooted tree. Thus, for example, the main species tree with proportion or weight (1-P(E)) is (E,((D,N),M)). The edge length E is going to be $(T_{EDNM} + 2Ne_{EDNM} \times G) \times \mu$, where $T_{EDNM}$ is the expected time that the ancestral population EDNM split into E and DNM, and $2Ne_{EDNM}$ is twice the effective ancestral population size, while G is the mean generation time in that ancestral population (e.g., Hudson 1992). This edge length is the same as the expected coalescent time of E and N (or E and D or E and M on this tree). On the other history, with proportion or weight P(E), the tree is ((E,D),(N,M)), and edge length E is equal to $(T(P(E)) + 2Ne_E{*}G) \times \mu$, where T(P(E)) is the age of the genetic injection from lineage E to lineage D. On this second tree edge length D equals edge length E. In this way it is possible to reconstruct the expected weighted tree for genes following either of the two possible "species" histories. The informative split weights that appear in the planar diagram for this model, which is what we really want, are just the internal edge lengths of these two trees.

I mentioned that it is the internal edge lengths of these two "species" trees that I really want. This is because we do not want to trust the external edge lengths, as explained in Waddell et al. (2011), because these are heavily contaminated with unknown amounts of "singleton" sequencing errors. The internal edge lengths and splits are expected to be much more robust. The second point to clarify here is that these two weighted trees have a one to one mapping to two different distance matrices. The spectrum of edge lengths (and hence the weighted sum of the two distance matrices) will map perfectly onto a planar graph since all splits can be described at once on a single circular splits system. As a reticulate diagram gets more complicated, then this may no longer be the case, and a planar diagram alone may be insufficient to untangle the splits.

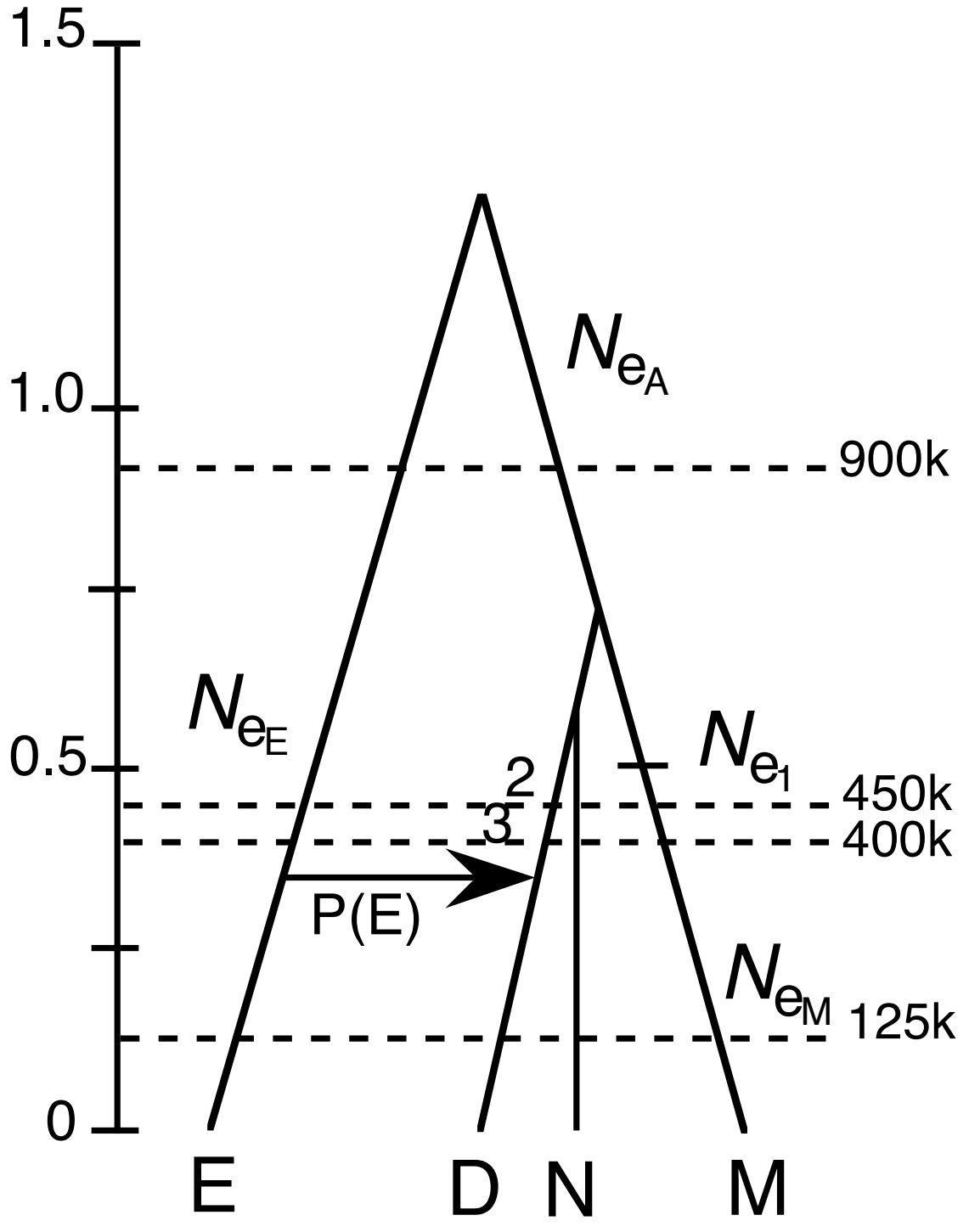

Figure 7. A plausible model of the reticulate interbreeding of the Denisovan lineage with a *Homo erectus*-like creature (the "archaic lecherous mans grandfather hypothesis" of Waddell et al. 2011, 2012). In this model, there is a rooted clock like-"tree of the expected mean coalesce times of four *Homo* genomes, which are, *Homo erectus*-like (E), *Homo denisova* (D), *Homo neanderthalensis* (N) and *Homo sapiens* (M). (The left scale is in ages of millions of years). The notch on the edge leading to moderns indicates the approximate mean coalescent time of San-Bushmen and all other modern human genomes. The dashed lines indicate, in order from deepest to most recent, the times of actual population divergences as the tree is traversed downwards and leftwards from the root. That is, $T_{EDNM}$ = 900k, $T_{DNM}$ = 450k, $T_{DN}$ = 400k, with the last line at 125k indicating when modern human populations began to probably began to diverge in thousands of years before present. The arrow indicates that a proportion, P(E), of the *erectus*-like genome was injected into the Denisovan lineage, at a time $T_{P(E)}$ of about 350,000 years ago. At each population split in the tree, there is an effective long-term population size, with $Ne_A$ = $Ne_{EDNM}$ being the ultimate ancestral population size, $Ne_E$ is the population size of lineage E at $T_{P(E)}$, $Ne_1$ = $Ne_M$ is the effective population size of the last ancestors of all modern humans, while 2 stands for $Ne_{DNM}$, and 3 indicates $Ne_{DN}$). These parameters are sufficient to describe what the relative informative split spectrum of genomic distances is expected to be under the infinite sites model.

What is needed now is to flesh out feasible parameter values to plug into the splits calculation. First, I will assume that the evolutionary calibration points of Waddell and Penny (1996) are about correct and that evolutionary rate of the great apes including humans is near uniform. Further, a modern human genomic rate of only ½ the evolutionary rate is problematic. If there really was a major change in substitution rate, then it would appear this needed to have occurred only in the last million years or so, since edge lengths to human and chimp on genomic trees are less than 10% different. Since the Denisovan and Neanderthal lineages are close in substitution rate to modern humans, this suggests any slow down was more than ~500,000 years ago. However, if it was a slowdown by a factor of 2, then it cannot be much older than twice this, since such a strong slowdown, for example, 2 million years ago, would alter the mean terminal genomic edge of human versus chimp by >90%, and this is not the case. Thus, if there was a single sharp genomic slowdown, it would appear to most likely affect the part of our model prior to 500,000 years ago but less than 2 million years ago. However, further arguing against any slowdown is the fact that the mtDNA of all *Homo* species so far, including the divergent mtDNA of Denisovans that looks like it came from an "erectus-like" creature at least 800,000 years ago, is clock-like.

Now we need to set up the divergence time and ancestral population size parameter values. The rooted weighted autosomal genome tree of the D, N and M lineages is approximately that seen in figure 1 of Waddell et al. (2011) (see also the NJ tree of Reich et al. 2010, although its edge lengths will differ slightly). The relative divergence times of the E versus DNM lineage and the DN versus M lineage are extrapolated from the tree in Krause et al. (2010) (assuming the mtDNA tracks the population tree closely, except it came across to D in gene injection P(E). The mtDNA splitting times of that tree are estimated as approximately 1mya and 0.5mya, respectively.

Long-term effective ancestral population sizes are drawn from a number of sources. The ancestral modern human effective population size seems to be somewhere in the range of 5,000 to 10,000 individuals, which, assuming a mean generation time of 25 years, translates to an autosomal pairwise coalescence time of ~250-500k (average 375k) for pairs of individuals in that population. The genomic divergence time indicated by the notch on the diagram in figure 7 at around 500kya is the approximate divergence time of San Bushmen genomes to non-African modern humans. This 500kya is made up of the 375,000ky plus a divergence time for Bushmen from non-African modern humans of around 125,000 years or about double the out of Africa event (a long assumed number largely corroborated by allelic data, mtDNA and now genomic sequences). The mtDNA results particular have remained fairly consistent on this general estimate over the years, and it fits increasingly well with archeology indicating modern behavior in southern-Africa prior to 70,000 years ago. Thus, we consider $Ne_M$ most feasible at around 7500, with a plausible range from about 3,000 to 12,000 assumed.

Estimating the other ancestral population sizes is now the challenge. Perhaps the best current guide as to what these were comes from figure 5b of Meyer et al. (2012), using a simple pairwise divergence model moving across the genome to estimate the date of coalescence and hence ancestral population size at different times in the past (Li and Durbin 2011). Such a method is not very good at differentiating a sudden bottleneck from a gradual decline (as seen in the simulation of figure of Meyer et al. 2012, figure S33), a feature shared by other similar methods of ancestral rate estimation (Kitazoe et al. 2007, figure 4). What Meyer et al. (2012) suggests is that the effective long term population size when D split from M was near its low over the past couple of million years or so. It appears to have been around 2/3 that of the size for modern humans and also only 2/3 that of what it was twice as long ago again (which in our scenario would be about 1 million years ago). Thus, we assume the $Ne_{DNM}$ parameter in figure 7 was 2/3 of 7,500 = 5,000 with a range of approximately 2/3 of 3000 = 2000 to 2/3 of 12,000 = 8000. From the same source, $Ne_{EDNM}$ was probably close to that of modern humans, again with a similar range of uncertainty. These three estimates are highly correlated.

It is also clear that the ancestral population size going out of Africa caused a major drop in modern long-term effective population size, and more particularly, this is very evident in figure 5B of Meyer et al. (2012) for the archaic Denisovan. What is also clear is that the Denisovan lineage appears to have had a genomic history most like the Karitiana tribe amongst the sample of modern humans, although stretched over a period of time about an order of magnitude longer. Thus it seems there was probably a pronounced out of Africa effect on the Denisovan lineage and, logically, on the DN ancestral lineage. We have a moderately stable estimate of the length of the pairwise genomic edge DN versus the edge leading to all moderns and it is about 3/4 for the site pattern ML calculations (e.g., Waddell et al. 2011, table 8 the ratio of g5/g4) or about 6/7 from the pairwise distance calculations such as figure 6 above. On the DN lineage, a similar amount of coalescence to the edge leading to modern humans seems to have occurred in about 50,000 years or less. Interestingly, the modern human analogue of the DN population squeeze in moving out of Africa appears to be almost the same size in figure 5B of Meyer et al. (2012) (despite modern humans apparently better exploiting the environment and keeping population size higher in general). We know the modern human out of Africa edge is almost certainly less than 50,000 years in duration, which suggests that the Denisovan and the Neanderthal lineages probably parted ways fairly soon after their own out of Africa. Thus the effective population size at $Ne_{DN}$ seems likely to have been less than ½ that of $Ne_{M}$. The effective population size on the Denisovan lineage prior to the injection of *erectus* genes is expected to have been even smaller given the trajectory see in figure 5 of Meyer et al. (2012). We do not need $Ne_{D}$ for the calculations below, and we can estimate $Ne_{DN}$ directly.

As an aside, a very interesting inference can be made tying the arguments of the previous paragraphs together. That is, there was almost certainly no out of Africa then back to Africa event of the last two million years or so in our own lineage. If, for example, large brained Asian erectus descendants had moved back into Africa and taken over (that is >80% genetic replacement), then this would suggest a likely triple bottleneck in the lineage leading to all modern humans, and nothing like this shows up on the plot of figure 5B of Meyer et al. This is a major point and supports the argument that Africa was center stage right through the evolution of *Homo*, not just *Australopithecus* and associates. That the current methods would be sensitive enough to detect such a series of bottlenecks remains to be fully assessed, but the maintenance of an effective population size leading to us of > 50% of the value at around 2mya (or around four times the age of the DN split from moderns) certainly contributes to the case for a lack a genetic replacement back migration to Africa from Asia.

That leaves the question of what to put for $Ne_{E}$ (actually, for immediate purposes, this is a superfluous parameter and does not affect the splits we need to calculate to estimate P(E)). The best proxy for $Ne_{E}$ would seem to be the terminal diversity seen in the Denisovan or Neanderthal genomes, which are much better understood, and were Eurasian pre-moderns. The Denisovan's seem to have leveled out at an effective population size of ~1/4 to 1/8 of that seen in $Ne_{M}$. The effective population size of Neanderthals also seems to be low, based on their mtDNA diversity being markedly less than that of modern humans. Indeed, across the whole of Europe, Neanderthal diversity is extremely low, with the only real diversity coming when a population in Mezmaiskaya is encountered, which puts the total diversity at about ½ that of modern humans, but with the suggestion of a huge degree of homogeneity over areas the size of Europe which is not typical of modern humans. Further, recent work presented at symposia by Svante Paabo in early 2013, suggests that Denisovan mtDNA may show a similar pattern, with two very similar copies and a third one that is nearly as divergent as that of San from other modern humans. This does not agree with the nuclear genome, so it might suggest the occasional, possibly forcible, capture of females from long isolated lineages. Such a behavior is now well characterized in chimps, and tends to involve a pattern typical of human warfare in earlier times. That is, groups of mature males on secretive raiding parties with the explicit purpose of murder (all males, and non-reproductive aged females), rape (usually just the females, but occasionally males when particularly aggressive) and abduction (of the reproductively attractive females, the one thing not so recently practiced in warfare). All usually directed at the closest distinct neighbors!

Thus we want to estimate the unknown parameter P(E) given our estimates of $Ne_{M}$ = 7,500, $Ne_{DNM}$ = 5,000, $Ne_{DM}$ to be estimated, $Ne_{EDNM}$ = 7,500, $T_{M}$ = 125kya, $T_{DN}$ = 400kya, $T_{DNM}$ = 450kya, $T_{EDNM}$ = 900kya and $T_{DE}$ of 350kya. The observed data are the informative edge lengths $E_{DN}$ = 39.7, $E_{M}$ = 47.6 and $E_{CD/NM}$ = 2.96 from the distance-based planar diagram of figure 1 in Waddell et al. (2011). With 3 pieces of information and using just their relative sizes, we do not need to know $\mu$, and can still solve for 2 free parameters. Doing this we get a perfect fit for P(E) = 0.0216 and $Ne_{DN}$ = 2,590. An example of the solution from the spreadsheet is given in figure 8. If we switch the data to the estimates from fig 6 when P is removed, we have E(DN) =

39.24, E(M) = 48.46 and E(CD) = 3.03, and very similar solutions of P(E) = 0.0217 and Ne(DN) = 2,689. Thus, in this scenario, despite edge CD always having a larger weight than the split NFHP, P(E) tends to be smaller than P(N) (which is in the range of 0.02 to 0.05 in Waddell et al. 2011, and similar estimates in Reich et al. 2011), since the erectus-like alleles being injected into D are even more distinct than the Neanderthal alleles.

The way to yield a larger percentage of ancestry (as opposed to derived alleles) in the Denisovan from an erectus-like form is to reduce the internode length of the genomic "species" trees contributing to splits other than ND and M. That is, increase the weight on the CD/NM split in figure 7 relative to the CM/DN and M/CDN splits (here M is the split weight leading to multiple modern genomes).

In terms of a sensitivity analysis, holding $\mu$ constant (which may need to be revisited in future), first alter G, the average generation time (again, being held constant across this analysis, which may need to be revised in future, with shorter times nearer the root). If G is increased to 30, then the solution is P(E):0.0181 and $Ne_{ND}$:3,356, while if it is decreased to 20, the solution is P(E):0.0254 and $Ne_{ND}$:1,436. Thus not a large change in P(E) but a more marked change in estimates of $Ne_{ND}$. With all other parameters at their best prior estimates, $Ne_{M}$ of 12,000 gives an untenable solution, but at 10,000 the solutions are tenable with P(E):0.0081 and $Ne_{NM}$:4,739. Dropping $Ne_{M}$ to 5,000 the solutions are P(E): 0.0351 and $Ne_{ND}$: 381, and for $Ne_{M}$ below this, the solution becomes untenable with $Ne_{ND}$ going to zero. Parameters P(E) and $Ne_{M}$ are strongly negatively correlated and sensitive to each other.

However, parameters $Ne_{EDNM}$, $Ne_{DNM}$ and $Ne_{M}$ are all positively correlated when based on the results shown in figure 3 of Meyer et al. (2012). Fixing their ratios to 1:2/3:1, then the solution for $Ne_{M}$:12,000 becomes P(E):0.0119 and $Ne_{DN}$:6,890. Likewise, $Ne_{M}$:5,000 gives solutions of P(E):0.0282 and $Ne_{DN}$:185, while lower values become untenable. Since these solutions are insensitive to $Ne_{E}$, then with all other parameters at their most likely assumed values and assuming this positive correlation, it seems likely that P(E) was in the range 0.012 to 0.28, with some bias against the low side, as it seems likely that $Ne_{DN}$ was markedly lower than $Ne_{M}$ and $Ne_{DNM}$, which is not the case for these solutions when P(E) is particularly low.

Sensitivity analyses on the assumed dates of species splits are also possible. As with $Ne_{E}$, this analysis is insensitive to $T_{P(E)}$. The biggest assumption is that the mtDNA tree accurately follows the species splitting times, which includes the assumption that the mtDNA came into D at the same time and from the same source as the autosomal loci. Putting that aside, but putting faith in the fact that mtDNA is a loci under strong negative selection, so its coalescent time should be <Ne/4, then a few percent deviation is expected on the relative times of the two critical splits. One of the greatest absolute uncertainties comes from the assumed time of the Human/Chimp divergence, which in turn hinges on the reliability of the estimates of the orangutan divergence time. With many analyses now adopting the results of Waddell and Penny (1996), setting this date at 16mya +/- 1mya, with an approximately normal error distribution, any major deviation from these expectations would propagate the effect (in percentage terms) to the inferred divergence times of the mtDNA. Thus, for example, if the orangutan really split off at 19mya, that would be a 3/16*100% or nearly 20% difference. Conversely, if views such as those held earlier by authors such as Vincent Sarich are correct, and a divergence time of as young as 12mya is correct, then that would be a 25% decrease. Holding the population sizes and generation time at their best assumed estimates, and ignore the fluctuations in the relative divergence times of the mtDNA due to exact coalescent time and stochastic fluctuations due to finite sequence length, then the optimal solutions are P(E):0.0248 and Ne(DN):1,667 for orangutan 20% older than expected and P(E):0.0160 and Ne(DN):3,737 if orangutan divergence is really 12mya. Thus an older orangutan time favors larger P(E) and smaller $Ne_{DN}$. Note, the relative time duration for $T_{DN}$ inferred here is largely unknown, but will be inversely proportional to $Ne_{DN}$.

Overall, it would seem sensible to assume that P(E) was most probably in the range of 1 to 3% until these variables can be pinned down appreciably better, and until factors such as changing generation time and relative mutation rate can be better settled.

A further note is that if erectus-like alleles did indeed enter the Papuan genome, the percentage is expected to be very small. The split weight of archaic alleles to Papuan, could represent an infusion of genes from a lineage another 300,000 years more separate than the lineage into Denisovan, if they ultimately traced back to the same archaic lineage. Or, up to an additional 300,000 + 700,000+ years separation if they came from a true, isolated East Asian *Homo erectus* derived lineage. In either case, the suggestion is that the amount of interbreeding could be 50% or less of P(E) estimated above, thus probably a smaller fraction than what many Europeans and East Eurasians got from Neanderthals.

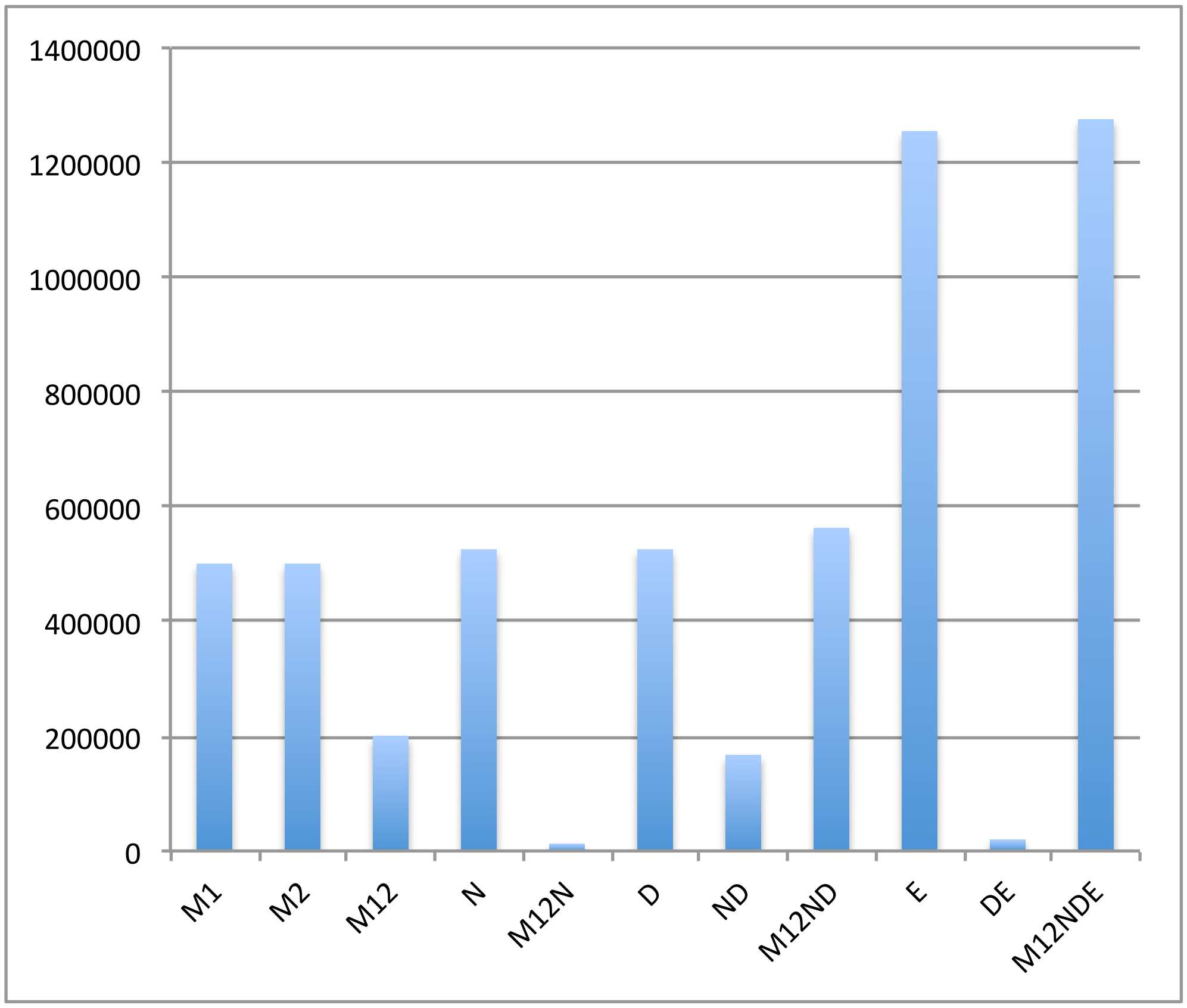

Figure 8. Example of the spread sheet output at the best solution with P(E) = 0.0216 and $Ne_{DN}$ = 2,590. Since we do not have genome E directly, signal DE becomes singletons for D, and we need to work exclusively with the informative splits M12, M12N and ND (here M12 indicates the edge leading to the two most divergence modern human lineages, while M12N is the same as the CD signal).

### 3.6 Quantitative assessment of the Phylogeny of *Homo*

The question now arises as to which archaic forms might have interbreed with the Denisovan lineage and also, which archaic form might have lead to such an apparently primitive genetic signature (CDNP) in modern Papuans?

Traditionally, the phylogenetic reconstruction of archaics has been based almost exclusively on discrete characters. That is, create a character matrix and then analyze it with parsimony. Both the reconstruction of the character matrix and the method of analysis present quite a few challenges. As for the data, it is often not clear what a character is. Metric quantities, such as the angle of this bone to that bone, are converted into discrete characters on a largely ad hoc basis. Other influential characters are often "seen" by some researchers, but not accepted by others. A good example is a "character" known as "an incipient suprainiac fossa." Many morphologists recognize that nearly all Neanderthal skulls show an distinct depression in the bone at the rear of the skull called a suprainiac fossa. What is much less clear is which, if any, other archaic skulls show a homologous feature. Some, will argue strongly that they can see an incipient suprainiac fossa in a fossil skull called SH5, yet many others doubt it is clear and/or homologous to the suprainiac fossa of Neanderthals. When this character is coded, it can be sufficient to swing the balance in favor of SH5 forming a clade with Neanderthals. However, problems with such a clade include a wide range of evidence suggesting that the modern/Neanderthal split was around 300 to 500,000 years ago (e.g. Krause et al. 2010), while the SH5 skull has been dated to around 600,000 years old with arguments for a minimum of 530kya (Bischoff et al. 2007).

One of the most promising ways to break out of this conundrum, and at least add more objective data, is quantitative skull shape evolution. A series of well-identified features are located in 3D space. From that point, a useful procedure is to produce pairwise minimum Euclidean distances between pairs of skulls. This can be done analytically using a closed form "Procrustes" procedure that is invariant to size (Bookstein 1991), so just shape differences are registered. In this case an L1 distance matrix from the discrete characters in Mounier et al. (2011) was created, then scaled to the mean of the Procrustes distance matrix, then the two added together.

From this matrix of minimum pairwise distances the distance data are analyzed via a pipeline very similar to that used for the genetic distances as developed in Waddell et al. (2010). An optimal WLS+ P=0 model is fitted, the residuals examined, then residual resampling procedures used to quantify the robustness of different parts of the tree (Waddell and Azad 2008). Fit measures such as g%SD can be compared directly with other data sets as a gauge of the precision being achieved by the data/method, and it is also possible to evaluate if non-phylogenetic approaches such as PCA or MDS are inferior descriptions of the data.

Results are shown in figure 9 and 10. The general structure that remains after applying residual resampling and reconstructing edges with WLS+ P=0, then ignoring all edges with less than 70% residual resampling support is that of a partly resolved tree.

In figure 10 it is clear that near modern skulls are strongly clustered with 98% residual resampling support into a single clade. Within this clade there is little well-supported detail, although there is the hint that LH18 and Skull 5 may be earliest diverging forms. Forms such as LH18 and Irhoud 1 had been identified in earlier times as Neanderthal-like, although most would now place them much closer to modern humans.

Neanderthal skulls also cluster together quite closely in figure 10 and they too receive substantial support of around 97% as a single clade with little resolution amongst them. The long edge to the "Old Man" of La Chapelle-aux-Saints skull is somewhat surprising. Remeasurements of the same landmarks on a cast at the American Museum of Natural History (AMNH) New York, by the author and a resident researcher (Brian Shearer), did not show the La Chapelle skull to deviate strongly from other Neanderthal measurements. However, the measurements reported and used here were upon the original skull and the person that did the measurements (Aure´lien Mounier) reports back no apparent fault in them (pers. comm.).

Near the root of the tree there is little resolution. The skull K1813 is typically assigned to the species *Homo habilis* and in turn is typically regarded as the least derived of the skulls in this assortment, and along with the Dmanisi skulls, has the oldest established age (listed in Mounier et al, 2011). Skulls 3733 and 3883 are typically considered good examples of *Homo ergaster*, which might be described as the African lineage of *Homo erectus*. *Homo erectus*, in the strict sense, is the species designation assigned in 1891 by Eugène Dubois, based on the Javan Trinil 2 skull cap which is now estimated at around 1mya (Swisher et al. 2000). In figure 9 we have fossils from Java that are dated around 1-1.5mya (Sangarin 17) or as young as ~120 to 546 kya (for Ngandong 6 and 14). Until recently it looked more than likely that a million years to 1.5 million years could separate early erectus from late erectus fossils in Java. However, given the difficulties of dating, the youngest date for the Sangiran structure is now fixed at 776kya (based on Ar39/Ar40 dating) while Ngandong specimens are now dated at around 546kya by Ar39/Ar40 (Indriati et al. 2011), so the actual spread in time of these fossils may be much less than earlier assumed. This lineage, overall, seems the most likely known candidate for the ancestors of the "Hobbit"® or *Homo floresiensis* (Morwood et al. 2004). Finally, there are the Zhoukoudian, China skulls (III and XII), which are better known as "Peking Man." In both time and morphology, they conform fairly well to the Javan fossils seeming to be somewhere in the interval of 400 to 800kya.

Note, while there is no robust structure amongst the erectus-like fossils other than the clustering of the Zhoukoudian specimens (figure 10), there are some hints of structure in figure 9 that might hopefully elaborate further with more data (for example, more landmarks and semi-landmarks). The promising hints are that most of the locality pairs, including Ngandong, Dmanisi, *Homo ergaster*, and Zhoukoudian do associate via adjacent splits in the original Neighbor Net planar diagram, so there seems to be some information on distinct genetic lineages preserved here.

Finally, perhaps the most exciting results in figure 10 are some quite robust features of distinct lineages of the taxonomic dumping ground of mid-Pleistocene humans, which have previously gone by the somewhat misleading name of archaic *Homo sapiens*. They are not *Homo sapiens* (that is, modern humans) and exactly what each of them is in the sense of relatedness to each other, or to better-understood and more recent lineages, remains highly controversial. Many of these fossils are single samples from a locality and have been given their own "species" designation. Most interestingly, some of these fossils with the lowest assigned minimum ages, listed in the range of approximately 125 to 250kya branch off around the modern /Neanderthal split (which seems fairly likely to have been around 400kya). Whether they predate that actual split or whether they associate with a particular lineage is unclear. However, the full planar network of figure 9 hints that the enigmatic European Steinheim skull at around 250kya might even be slightly closer to the modern human African lineage than the classic-Neanderthal European lineage that came to dominate Europe and push as far South as Sinai and as far East as Siberia before being swept aside by the modern out of Africa event.

The next substantial result is that the beautifully preserved SH5 skull from Spain, perhaps most reliably dated at near 600kya (Biscoff et al. 2007) has substantial support of 94% as a sister lineage to a clade of modern Humans, Neanderthals and a geographically widespread collection of fossil forms that may date from the period ~150 to 300kya as just mentioned (Jinniu-Shan from China, Steinheim from Europe and Kabwe from Africa). In this analysis it seems to form its own distinct lineage and might presently be the closest sample we have to our own ancestors from that time period of ~500 to 700 kya. It is separated from earlier diverging lineages with fairly good confidence of around 88% residual resampling support.

Sister to all the forms described so far are the well preserved Petralona (Greece) skull, listed here as in Mounier et al. at 200kya or older, but quite probably closer to 400kya and also the Dali skull from central China of at least 280 kya. Whether these represent one lineage or two separate lineages is unclear. There is, then, moderate 75% support for all the afore mentioned specimens forming a clade, with the erectus-type forms all clearly distinct and earlier diverging. There is one further particularly interesting result here, and that is that the Ceprano skull might represent a late surviving lineage that split from our ancestors perhaps near 1mya ago. The planar diagram hints at this, but the residual resampling support is less than 70% in these analyses, so that it is not clearly differentiated from the more classical erectus-like forms.

As a note, while it has been popular recently to call lots of middle Pleistocene forms *Homo heidelbergensis*. However, such a term seems little better a solution for the present state of affairs than calling them all archaic *Homo sapiens*. Part of the problem stems from the fact that the holotype of *H. heidelbergensis* is a robust jaw that, in some sense, might as well have come from Herman Munster (biogeographically also originally from that part of the world). That is to say, without an associated skull, defining *Homo heidelbergensis* morphologically is difficult, and often seems to revert nearly as much to biogeography and age as much as to morphology. The results in figure 10 hint at considerable robust structure and multiple lineages that might have been evolving and dispersing in the middle Pleistocene.

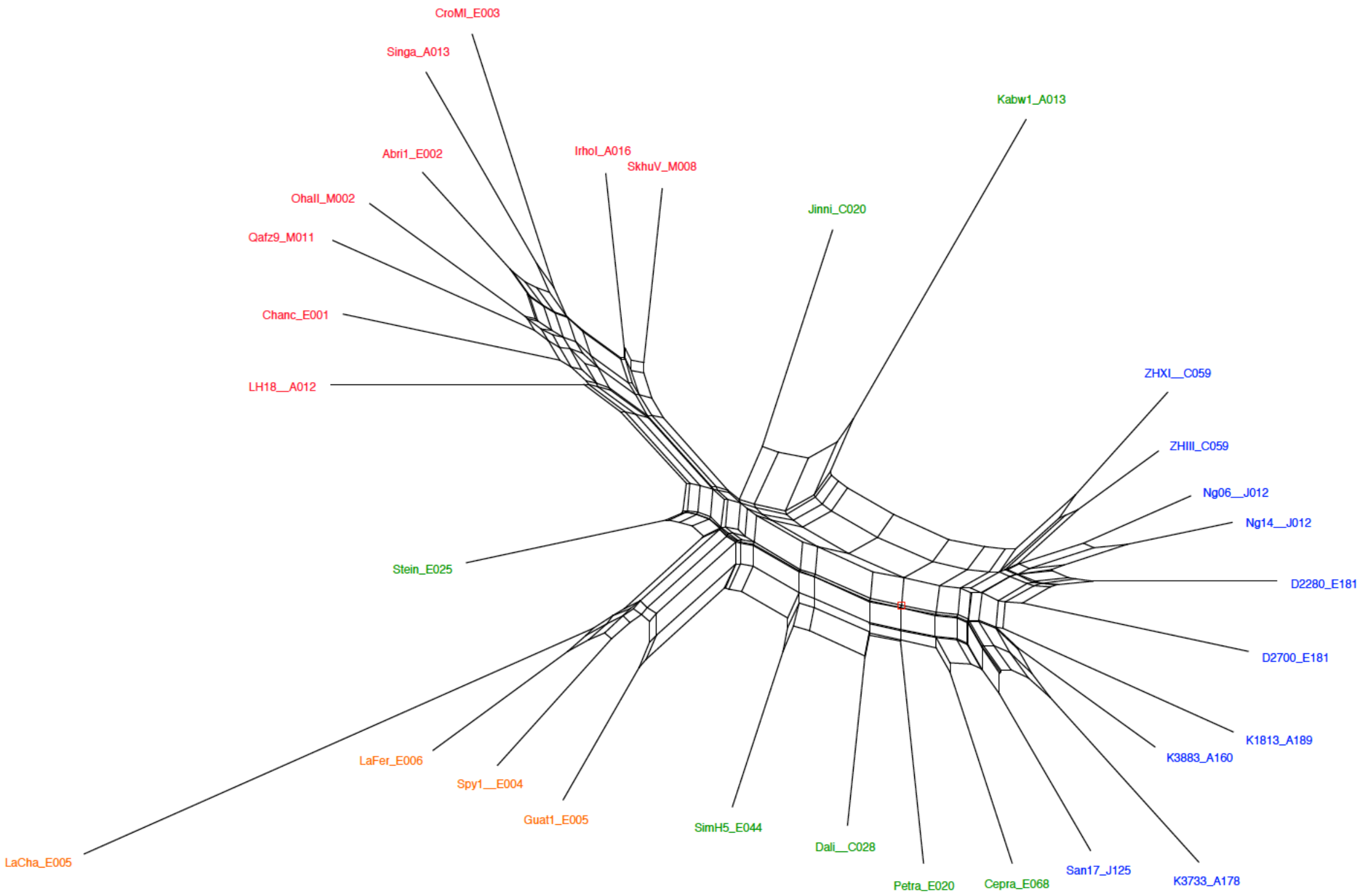

Figure 9. Planar diagram of relationships of fossil *Homo* skulls based on using Neighbor Net followed by OLS+ P = 0 to estimate edge weights. Based on a combined matrix of pairwise “Procrustes” minimum pairwise Euclidean distances between 14 landmarks measured on a wide range of fossil skulls and an L1 distance metric on discrete characters (data from Mounier et al. 2011, with 3D landmark measurements kindly provided by Aure´lien Mounier). (Figure from Colloquia given by the author at AMNH/NYCEP, Harvard Medical School and Yale Medical School, April 2013). At the tips of the edges are region of origin (A: Africa, E; Europe, J: Java and C: China) along with approximate minimum ages based on Mounier et al. (2011). Note, for some of these skulls plausible minimum ages may be gross under estimates of the true age. In particular, for our discussion, the Ngandong skulls are probably anywhere from ~200 to 600 kya (Indriati et al. 2011), while SH5 is probably older than 500kya (Bischoff et al. 2007). Cepra (Ceprano) might even be as young as 350kya (Muttoni et al. 2009), suggesting multiple lineages around Europe in that general time period. Colors are red for near modern morphology, orange for Neanderthal, green for mid-Pleistocene forms without confident assignment, and blue for what are commonly regarded as *Homo habilis/ergaster/erectus* forms that also act as the outgroup.

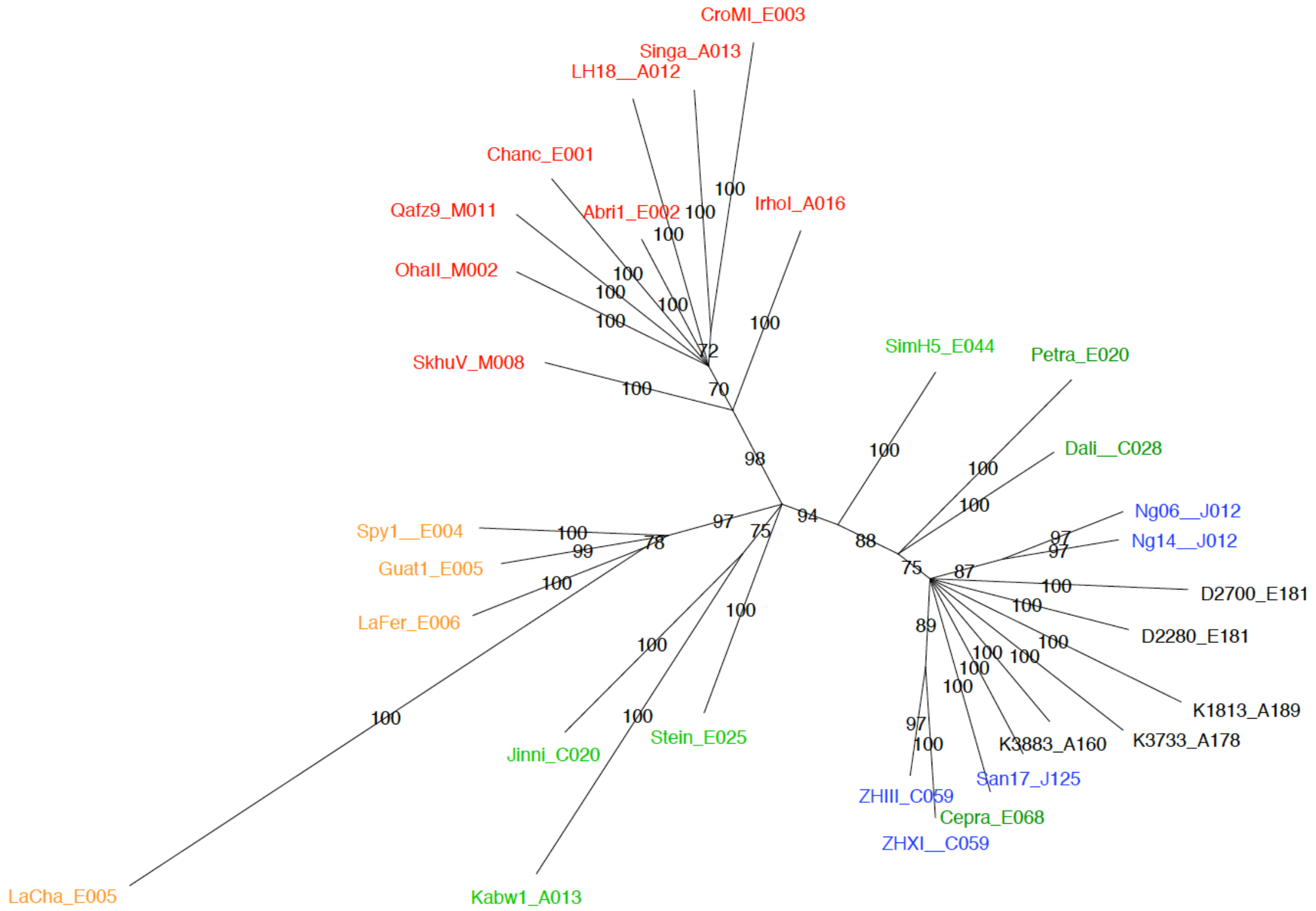

Figure 10. Residual resampling analysis (via normal approximation) of the distance data used in figure 9. Each replicate data set was analyzed with OLS+ P=0. The percentage at which an edge had a weight > 0 is noted. Any split with a resampling proportion of less than 70% was discarded.

## 4 Discussion

The new results presented here generally corroborate and reinforce the prior hypotheses of archaic interbreeding. In terms of specific splits, the most weighty non-tree split that emergences from this data via the analysis of pairwise distances and planar graphs is the enigmatic Chimp+Denisova+Neanderthal+Papuan split. Also well supported and very stable when removing taxa is the CD split for archaic genes going into Denisovan (Waddell et al. 2011). In comparison to some highly publicized splits such as the European+East Asian split (here French+Han) these splits have much more weight and comparable residual resampling support. Both have far more weight and support than other a priori splits such as Denisova+Han (Skoglund and Jacobsson 2011), Neanderthal+modern-out-of-Africa (Green et al. 2010), more Neanderthal into East rather than West Eurasians (e.g., Meyer et al. 2012), or of archaic genes into San, Chimp+San (Hammer et al. 2012). Thus, the more complex archaic signals for Papuan and Denisovan, including an avoidance of NP and DP in favor of CD, DNP and CDNP remain large even as different taxa are removed.

In contrast to a lot of non-tree splits, the X chromosome seems to show a much more tree-like pattern of evolution. There are non-tree splits, such as Han+Karitiana, but they do not match exactly to the autosomal patterns in terms of exactly what they are or in their weight. In particular, there is no evidence of archaic X genes in modern humans. This corroborates the lecherous archaic man hypothesis of Waddell et al. (2011) of much less archaic material on X, but unless there were mutants producing only Y-bearing sperm, then such a strong lack of archaic X suggests that selection against the archaic X genes in modern human populations played the major role. This complete lack of evidence of archaic gene introgression into X is only evident in the Meyer et al. 2012 data; the 2010 Reich chromosome X data shows a reduced signal of a Denisova+Papuan split, which may be mismapped reads.

Novel analyses of morphological data provides robust evidence that the middle Pleistocene saw multiple lineages evolving and widely distributed across Eurasia. Considering the structure of figure 10 also allows more precise predictions matching genes to morphology. As to the morphology of the Denisovan individual, there seems a better chance that it was somehow allied with the Jinniu-Shan fossil (which looks a bit Neanderthal) than with the Dali fossil, which appears in figure 10 as a much deeper lineage. If a date for the Denisovan lineage separate to the Neanderthal lineage of less than 300kya becomes attractive, then the younger date for Jinniu-Shan might also push this to be a more attractive proposition.

In terms of what is Eurasian and pre-Neanderthal/human split, that could have contributed more archaic genes to the Denisovan (the Chimp+Denisovan split), then distinct possibilities include SH5, Petralona, Dali, Ceprano and, of course, the classic erectus. Given that Krause et al. (2010) show that the contributing lineage was probably less than 1 mya, then that tends to rule out the truly erectus-like and leave multiple possible lineages of middle Pleistocene forms (figure 10).

In terms of what might have contributed the apparent strongly archaic signal being picked up in Papuan, it might be a more intense version of the same pairing the Denisovan genome shows, only more archaic going in, or it could be truly from a more *Homo erectus* like, rather than middle Pleistocene like, source. The evidence seems to suggest a SE Asian erectus lineage almost certainly extant to 500kya, but proving survival later than that needs improved dating. However, the Hobbit® (*Homo floresiensis*) was probably *Homo erectus* derived and thus Papuans could conceivably have received a few genes via that lineage, or even via a relative of the Denisovan lineage, enriched in archaic genes from the last of the Ngandong forms.

As a caution, it is useful to point out, that unlike the CD split, the CDNP split has yet to be clearly corroborated by a deficiency of derived alleles in Papuan, after the known archaic interbreedings are taken into account (Waddell et al. 2011). Thus, at least part of that signal strength could be an unexpectedly large general sharing of San+Yoruba+French+Han alleles after the divergence of Papuan, despite this not being supported by strong weights on more feasible pairings, such as Yoruba+French. However, in some of the analyses of Waddell et al. (2011, 2012), the pattern DNSYFH did appear in clear excess, which might be related to an erectus-like+Papuan signal, since it is hard to imagine how these other disparate lineages would exchange genes to the exclusion of Papuan. Higher quality site pattern data is needed to test this.

Finally, having estimated ~2% particularly archaic alleles into Denisovan, and a hint of a similar proportion into Papuan, perhaps a short note on what having archaic genes might mean is in order. While there are web sites that proclaim things like “erectus walks amongst us” that are often both intellectually weak and overtly racist, it is useful to consider more carefully what having Neanderthal, erectus or other archaic genes in us might mean. One inclination is to too often think of these genes as “inferior” or “lesser” in some way. It should be mentioned, however, that most of these introgressions appear to have been occasional injections of archaic genes into modern populations. As such, these “alternative” genes would have been under intense selection. The modern human genome seems to be a highly coadapted complex, including with regard to higher cognitive functions, and anything that comes in to disrupt that will be running into a very strong selective headwind. Indeed, due to linkage, many potentially useful gene variants may not have survived introgression due to disruptive nearby genetic variants. If erectus-like alleles are truly present in Papuans, then they should have been under similarly intense selective pressure to the many Neanderthal genes that all non-Africans have in a similar amount. As such, any erectus-like genes that survive to the present have been under selection for a relatively long time. The average expected block size from introgression 50,000 years ago is now down into chunks of genome approximately 10 kb. It shall be very interesting to see if they, like the Denisovan and Neanderthal-like alleles that Abi-Rached et al. (2011) consider, may have further assisted Oceanian populations to adapt to a variety of new conditions. It might even be possible that a few highly advantageous erectus derived variants might have spread back into mainland Eurasia.

## Acknowledgements

This work was supported by the National Institutes of Health [ 5R01LM008626 to P. J. W.].

Thanks to David Bryant, Dick Hudson, Yunsung Kim, Martin Kircher, Hiro Kishino, Aure´lien Mounier, Nick Patterson, Phillip Rightmere, Brian Shearer, Dave Swofford, and Xi Tan for helpful discussions and/or assistance with data/software. Thanks to Eric Delson, David Reich and Kenneth Kidd for invitations to present earlier versions of this work at colloquia.

## Appendix

Data used from Reich et al. (2010). First is the table of site patterns (modified from the data communicated by Nick Patterson) and second the distance matrix calculated from these using PAUP*.

Table A1. Data as site pattern frequencies used in the analyses of Reich et al. (2010) as communicated by Nick Patterson and reordered by the author.

| Pattern | auto | chrX | Pattern | auto | chrX | Pattern | auto | chrX | Pattern | auto | chrX |
|---|---|---|---|---|---|---|---|---|---|---|---|
| Identity | 334171159 | 5222564 | N | 92072 | 1356 | D | 57327 | 825 | DN | 11849 | 186 |
| P | 98756 | 1517 | NP | 1172 | 7 | DP | 1071 | 6 | DNP | 1116 | 9 |
| H | 97234 | 1373 | NH | 1213 | 7 | DH | 639 | 10 | DNH | 853 | 7 |
| HP | 5416 | 74 | NHP | 413 | 4 | DHP | 395 | 4 | DNHP | 555 | 8 |
| F | 85418 | 1243 | NF | 1165 | 14 | DF | 613 | 5 | DNF | 935 | 21 |
| FP | 4601 | 59 | NFP | 347 | 2 | DFP | 337 | 6 | DNFP | 483 | 9 |
| FH | 5127 | 61 | NFH | 429 | 10 | DFH | 321 | 4 | DNFH | 465 | 15 |
| FHP | 5340 | 99 | NFHP | 559 | 5 | DFHP | 513 | 6 | DNFHP | 990 | 21 |
| Y | 89751 | 1287 | NY | 1131 | 16 | DY | 1097 | 14 | DNY | 1275 | 28 |
| YP | 2523 | 42 | NYP | 314 | 3 | DYP | 301 | 2 | DNYP | 423 | 9 |
| YH | 2912 | 49 | NYH | 317 | 1 | DYH | 261 | 3 | DNYH | 412 | 6 |
| YHP | 2002 | 37 | NYHP | 309 | 4 | DYHP | 294 | 3 | DNYHP | 511 | 10 |
| YF | 3141 | 53 | NYF | 331 | 6 | DYF | 305 | 2 | DNYF | 468 | 14 |
| YFP | 1777 | 45 | NYFP | 267 | 6 | DYFP | 245 | 1 | DNYFP | 413 | 8 |
| YFH | 2064 | 37 | NYFH | 320 | 7 | DYFH | 241 | 1 | DNYFH | 466 | 6 |
| YFHP | 4234 | 93 | NYFHP | 931 | 12 | DYFHP | 866 | 8 | DNYFHP | 2456 | 33 |
| S | 81426 | 1226 | NS | 1419 | 16 | DS | 1381 | 14 | DNS | 1666 | 24 |
| SP | 1966 | 28 | NSP | 294 | 3 | DSP | 293 | 4 | DNSP | 486 | 16 |
| SH | 2136 | 35 | NSH | 337 | 3 | DSH | 276 | 3 | DNSH | 505 | 9 |
| SHP | 1478 | 29 | NSHP | 237 | 0 | DSHP | 218 | 2 | DNSHP | 500 | 7 |
| SF | 2432 | 41 | NSF | 350 | 6 | DSF | 286 | 2 | DNSF | 484 | 19 |
| SFP | 1210 | 27 | NSFP | 204 | 2 | DSFP | 235 | 2 | DNSFP | 425 | 5 |
| SFH | 1436 | 28 | NSFH | 325 | 2 | DSFH | 237 | 2 | DNSFH | 402 | 9 |
| SFHP | 2571 | 51 | NSFHP | 684 | 5 | DSFHP | 650 | 7 | DNSFHP | 1866 | 23 |
| SY | 3862 | 46 | NSY | 565 | 12 | DSY | 564 | 4 | DNSY | 898 | 14 |
| SYP | 1014 | 24 | NSYP | 242 | 2 | DSYP | 278 | 3 | DNSYP | 570 | 13 |
| SYH | 1137 | 24 | NSYH | 239 | 1 | DSYH | 263 | 1 | DNSYH | 584 | 5 |
| SYHP | 1366 | 29 | NSYHP | 389 | 3 | DSYHP | 398 | 0 | DNSYHP | 1120 | 12 |
| SYF | 1299 | 45 | NSYF | 302 | 4 | DSYF | 286 | 1 | DNSYF | 587 | 11 |
| SYFP | 1204 | 33 | NSYFP | 359 | 5 | DSYFP | 399 | 4 | DNSYFP | 1098 | 16 |
| SYFH | 1643 | 20 | NSYFH | 502 | 9 | DSYFH | 423 | 2 | DNSYFH | 1179 | 14 |
| SYFHP | 7891 | 146 | NSYFHP | 4776 | 45 | DSYFHP | 4069 | 63 | DNSYFHP | 902013 | 10885 |

Table A2: Pairwise autosomal p-distances based upon sites from Reich et al. 2010, which pass the filters and at which all eight taxa are called.

| | | | | | | | | |
|---|---|---|---|---|---|---|---|---|
| Chimp | 0 | 3021.76 | 3133.74 | 3131.33 | 3172.91 | 3184.99 | 3223.37 | 3221.96 |
| Denis | 3021.76 | 0 | 562.47 | 640.19 | 681.01 | 693.23 | 730.6 | 725.21 |
| Neander | 3133.74 | 562.47 | 0 | 746.25 | 787.09 | 794.49 | 831.13 | 831.73 |
| San | 3131.33 | 640.19 | 746.25 | 0 | 690.22 | 700.58 | 738.88 | 745.27 |
| Yoruba | 3172.91 | 681.01 | 787.09 | 690.22 | 0 | 714.09 | 752.34 | 759.99 |
| French | 3184.99 | 693.23 | 794.49 | 700.58 | 714.09 | 0 | 708.71 | 718.58 |
| Han | 3223.37 | 730.6 | 831.13 | 738.88 | 752.34 | 708.71 | 0 | 745.11 |
| Papuan | 3221.96 | 725.21 | 831.73 | 745.27 | 759.99 | 718.58 | 745.11 | 0 |

Table A3: Pairwise chromosome X p-distances based upon sites from Reich et al. 2010, which pass the filters and at which all eight taxa are called.

| | | | | | | | | |
|---|---|---|---|---|---|---|---|---|
| Chimp | 0 | 2380.77 | 2488.31 | 2506.61 | 2543.98 | 2563.81 | 2572.59 | 2601.95 |
| Denis | 2380.77 | 0 | 494.23 | 617.79 | 655.16 | 663.93 | 686.43 | 709.69 |
| Neander | 2488.31 | 494.23 | 0 | 723.8 | 753.55 | 762.32 | 794.35 | 822.19 |
| San | 2506.61 | 617.79 | 723.8 | 0 | 666.6 | 671.94 | 703.59 | 725.33 |
| Yoruba | 2543.98 | 655.16 | 753.55 | 666.6 | 0 | 670.79 | 704.73 | 722.66 |
| French | 2563.81 | 663.93 | 762.32 | 671.94 | 670.79 | 0 | 663.55 | 691.77 |
| Han | 2572.59 | 686.43 | 794.35 | 703.59 | 704.73 | 663.55 | 0 | 702.06 |
| Papuan | 2601.95 | 709.69 | 822.19 | 725.33 | 722.66 | 691.77 | 702.06 | 0 |